\documentclass[lettersize,journal]{IEEEtran}

\usepackage{soul}
\usepackage{epic,eepic,units}
\usepackage{psfrag}
\usepackage{latexsym}
\usepackage{longtable}
\usepackage{mathrsfs}
\usepackage{bigstrut}
\usepackage{graphicx}
\usepackage{pifont}
\usepackage{amsmath}
\usepackage{setspace}
\usepackage{subfigure}
\usepackage{mathtools}
\usepackage{mathcomp}
\usepackage{mathdots}
\usepackage{rotating}
\usepackage{array}
\usepackage{multirow}
\usepackage{multicol}
\usepackage{color}
\usepackage{algorithm}
\usepackage{algorithmic}

\usepackage{cite,amssymb}
\usepackage[dvipsnames]{xcolor}

\usepackage{euscript,yfonts,dsfont,bbm,amstext,wasysym,pdfsync,amsfonts,arydshln,framed}

\usepackage{tikz}
\usetikzlibrary{arrows,shapes,trees,calc,positioning}
\usetikzlibrary{matrix}

\newtheorem{theorem}{Theorem}

\newcommand{\enma}[1]   {\ensuremath{#1}}

\newcommand{\beq}{\begin{equation}}
\newcommand{\eeq}{\end{equation}}
\newcommand{\bseq}{\begin{subequations}}
\newcommand{\eseq}{\end{subequations}}
\newcommand{\beqn}{\begin{eqnarray}}
\newcommand{\eeqn}{\end{eqnarray}}
\newcommand{\ba}{\begin{array}}
\newcommand{\ea}{\end{array}}
\newcommand{\bct}{\begin{center}}
\newcommand{\ect}{\end{center}}
\newcommand{\btmz}{\begin{itemize}}
\newcommand{\etmz}{\end{itemize}}
\newcommand{\benum}{\begin{enumerate}}
\newcommand{\eenum}{\end{enumerate}}

\newcommand{\cH}{\enma{\mathcal H}}
\newcommand{\cL}{\enma{\mathcal L}}

\newcommand{\norm}[1]{\| #1 \|}                 

\newcommand{\w}{\omega}

\newcommand{\diag}      {\enma{\mathrm{diag}}}

\newcommand{\trace}     {\enma{\mathrm{trace}}}

\newcommand{\col}       {\enma{\mathrm{col}}}

\newcommand{\inner}[2]{\left\langle #1,#2 \right\rangle}

\newcommand{\bv}{{\bf v}}

\newcommand{\matbegin}{
        \left[
}
\newcommand{\matend}{
        \right]
}

\newcommand{\tbt}[4]{
  \matbegin \begin{array}{cc}
       #1 & #2 \\ #3 & #4
       \end{array} \matend }

\newcommand{\be}{\begin{equation}}
\newcommand{\ee}{\end{equation}}

\newcommand{\cplxs}{ C\kern -.35em \rule{0.03 em}{.7 ex}~   }

\def\complex{\hbox{C\kern -.45em \rule{0.03 em}{1.5 ex}}~}

\newcommand{\bi}{\begin{itemize}}
\newcommand{\ei}{\end{itemize}}

\newcommand{\bu}{{\bf u}}

\newcommand{\bw}{{\bf w}}
\newcommand{\bbZ}{\mathbb{Z}}

\newcommand{\bbR}{\mathbb{R}}
\newcommand{\bbC}{\mathbb{C}}

\newcommand{\btab}{\begin{tabular}}
\newcommand{\etab}{\end{tabular}}

\newcommand{\bpsi}{\mbox{\boldmath$\psi$}}

\newcommand{\non}{\nonumber}
\newcommand{\mrd}{\mathrm{d}}
\newcommand{\mre}{\mathrm{e}}
\newcommand{\mri}{\mathrm{i}}

\newcommand{\ds}{\displaystyle}

\newcommand{\eps}{{\epsilon}}

\newcommand{\bA}{\mathbf{A}}
\newcommand{\bB}{\mathbf{B}}

\newcommand{\bk}{\mathbf{k}}
\newcommand{\DefinedAs}[0]{\mathrel{\mathop:}=}
\newcommand{\vsp}{\vspace*{0.15cm}}
\DeclareMathOperator*{\argmin}{arg\,min}

\DeclareMathOperator*{\minimize}{minimize}
\DeclareMathOperator*{\subject}{subject~to}

\definecolor{bgblue}{rgb}{0.04,0.19,0.53}
\definecolor{dblue1}{rgb}{0,0.3,0.7}
\definecolor{dred}{rgb}{0.4,0.2,0}

\newcommand{\mrn}{{\mathrm{n}}}
\newcommand{\mrm}{{\mathrm{m}}}

\begin{document}


\title{\LARGE 
${\mathcal H}_2$-optimal riblet profiles for turbulence suppression in channel flows
}

\author{Daniel Hartman and Armin Zare
\thanks{This work was supported in part by the Air Force Office of Scientific Research under award FA9550-23-1-0219. D.\ W.\ Hartman and A.\ Zare are with the Department of Mechanical Engineering, University of Texas at Dallas, Richardson, TX 75080, USA. Emails: {\tt\small daniel.hartman@utdallas.edu}; {\tt\small armin.zare@utdallas.edu}.}
\thanks{D. Hartman,  and A.~Zare are with the Department of Mechanical Engineering, University of Texas at Dallas, Richardson, TX 75080, USA.\newline Email: {\tt\small armin.zare@utdallas.edu}}
}

\maketitle
\thispagestyle{empty}
\pagestyle{empty}

\begin{abstract}
The drag-reducing properties of streamwise-elongated surface corrugations, commonly referred to as riblets, have been extensively studied through experiments, simulations, and reduced-order modeling. Notably, the linearized Navier–Stokes framework introduced by Ran, Zare, and Jovanovic (J. Fluid Mech., vol. 906, 2021) employs volume penalization to model riblet-induced boundary conditions and predict drag reduction. These approaches typically quantify drag reduction as a function of a single geometric parameter, such as spanwise spacing or groove cross-sectional area, evaluated over fixed riblet shapes. As a result, identifying optimal riblet geometries across a broader design space often necessitates exhaustive parametric studies. To address this problem, we propose an optimization framework for the systematic design of optimal riblet shapes. Accordingly, we reformulate the volume penalization term in the linearized dynamics as a structured state-feedback operator. Targeting maximal suppression of turbulence, we pose a composite optimization problem that minimizes the steady-state variance of velocity fluctuations and riblet-induced dynamical augmentations while maintaining a constant bulk-flux. For a fixed spanwise wavelength (riblet spacing), this framework yields optimal wall-normal profiles that achieve maximal turbulence attenuation.
\end{abstract}


\section{Introduction}

The influence of engineered surface roughness on skin-friction drag has been studied for decades. Early work in this area was motivated by the distinctive texture of shark skin and its apparent role in facilitating efficient locomotion through water~\cite{reidin82,becrei85,berbarhoprei86}. These observations prompted further investigation into the physical mechanisms underlying this biological phenomenon~\cite{becbruhag00,fuyuabai17} and, subsequently, the development of biomimetic surfaces designed to reproduce its drag-reducing characteristics.

Among such surfaces, riblets---streamwise-aligned surface corrugations used for passive flow control---have received considerable attention. Their capacity to reduce turbulent skin-friction drag has been demonstrated analytically~\cite{becbar89}, numerically~\cite{garjim11a,garjim11b,elschuyoo07}, and experimentally~\cite{becbruhagvanjgthop97,sasmamiwamur14}. Nevertheless, identifying optimal riblet geometries and spacings remains challenging. A principal difficulty lies in accurately modeling turbulent flow over rough surfaces, while the resulting design problems are often nonconvex and computationally demanding. Consequently, advances in riblet design have historically relied heavily on improved physical understanding of the mechanisms responsible for drag reduction~\cite{wal82,wal83}, rather than on systematic optimization alone.

Most studies of riblet design focus on analyzing the flow physics associated with prescribed geometries rather than systematically optimizing the riblet shape~\cite{modendhutchu21,ranzarjovJFM21,garjim11a}. Several studies have nevertheless addressed the design of riblets for drag reduction. These approaches often rely on high-fidelity simulations, such as direct numerical or large-eddy simulations, which provide detailed flow information but impose substantial computational costs. For example, Lee et al.~\cite{leemoalimgumar24} combined Bayesian optimization with large-eddy simulations to develop a design-by-morphing framework. Friedmann~\cite{fri10}, by contrast, formulated an objective functional based on wall shear stress and used it to determine the optimal periodicity of sinusoidal surface roughness.

In this study, we develop a riblet shape-optimization framework that accommodates a broad class of geometries rather than restricting the design space to a single prescribed shape of riblets, e.g., triangular, blade, or scalloped riblets. Unlike approaches that rely exclusively on computationally intensive nonlinear flow simulations, our framework employs low-complexity models obtained by linearizing the flow dynamics about the long-time-averaged velocity field over riblets. This formulation enables systematic exploration of the riblet design space while retaining the influence of surface geometry on the underlying turbulent dynamics.

Linearized Navier-Stokes (NS) equations have been widely used to characterize the structural and statistical properties of turbulent wall-bounded shear flows~\cite{mcksha10,hwacosJFM10a,zarjovgeoJFM17,zargeojovARC20,jovARFM21,bharodberleozarEnergies23}. In this class of models, the governing equations are cast in evolution form and driven by deterministic or stochastic forcing to excite physically relevant flow responses. Such models have also enabled physical analysis and discovery~\cite{jovbamJFM05,hewzarPRF22,abozarJFM23}, in addition to the development of flow estimation~\cite{aboleorotzarJRSE25} and control strategies, including sensor-free techniques for turbulence mitigation based on surface-normal blowing and suction in the form of streamwise traveling waves~\cite{moajovJFM10,liemoajovJFM10}, transverse wall oscillations~\cite{moajovJFM12}, and riblets~\cite{ranzarjovJFM21,naszarACC25}.

Here, we adopt the modeling framework developed in~\cite{ranzarjovJFM21} to represent both the rough-wall boundary conditions and the statistical signatures of turbulent channel flow over riblets. Following Ref.~\cite{chaluh19}, this framework employs a volume-penalization technique~\cite{khamanhammoi10} to approximate the influence of a spatially periodic surface on the flow. Rather than deforming the computational mesh to conform to the rough geometry, volume penalization introduces a spatially varying, static feedback term into the momentum equations that suppresses the flow within the solid region and restricts the dynamics to the fluid domain surrounding the riblet surface. The framework further incorporates the turbulence-modeling approach of Moarref and Jovanovi\'c~\cite{moajovJFM12} to account for the influence of background turbulence on the mean velocity and, consequently, on skin-friction drag. Specifically, the turbulent eddy viscosity for channel flow over riblets is inferred from the second-order statistics of the eddy-viscosity-enhanced linearized NS equations. The resulting corrections to the turbulent viscosity are then used to determine the riblet-induced modifications to the mean flow and drag.

Using this linear low-complexity model, we formulate an unconventional optimal-control problem in which the state variance quantifies the energy of velocity fluctuations as a surrogate measure of drag~\cite{ranzarjovJFM21}, while the control effort quantifies the magnitude of the dynamical augmentation induced by the riblets. The optimization is subject to both structural and physical constraints. The structural constraints prescribe the sparsity pattern of the feedback gain matrix required to represent harmonic interactions among the mean flow, the volume-penalization field, and the velocity fluctuations. The physical constraints enforce geometric connectivity, consistency of the dynamical penalization, and preservation of a constant bulk flux when quantifying the influence of riblets on drag.

In unconstrained state-feedback design, an optimal stabilizing feedback gain can often be obtained from the solution of an algebraic Riccati equation. Imposing structural constraints on the gain matrix substantially complicates the problem and has motivated customized solution methods, including projected-gradient algorithms~\cite{linfarjovTAC11al}. More general formulations that impose additional restrictions directly on the feedback gain, such as linear equality constraints, remain an active area of research~\cite{chemalitomlee22,yuachezhayan23}.

We address the physical constraints in our formulation by parameterizing admissible solutions using a library of riblet geometries represented by signed-distance functions. These are level-set functions that encode the distance to a parameterized boundary curve; see, e.g.,~\cite{malsetvem02}. We show that linear combinations of these functions, embedded within the structured feedback matrix that augments the linearized dynamics, generate morphed riblet geometries that satisfy the prescribed shape constraints. This parameterization also yields a reformulated optimization problem that is amenable to gradient-based methods.

To solve the resulting problem, we develop an augmented-Lagrangian-based algorithm. Because the formulation is nonconvex, optimization trajectories initialized at nearby points may converge to distinct local minima with different fluctuation-energy and drag-reduction characteristics. We therefore employ an ensemble of randomly generated initial conditions selected to ensure that the $\cH_2$ objective and its gradients are well defined. For a prescribed riblet spacing and under constant-bulk-flux conditions, the optimized solutions reveal a recurring preference for scallop-shaped riblet geometries that effectively suppress turbulent fluctuations.

The main contributions of this paper are as follows:
\begin{enumerate}
  \item Formulation of riblet shape optimization as a state-feedback synthesis problem subject to application-specific structural and physical constraints, in which riblet geometry is represented through structured dynamical augmentations of the linearized NS equations.
  \item Development of a customized augmented-Lagrangian algorithm for solving the resulting constrained $\cH_2$ optimization problem.
  \item Demonstration that a level-set representation of riblet geometries based on signed-distance functions enforces physical constraints associated with geometric connectivity and consistent volume penalization.
\end{enumerate}

The remainder of the paper is organized as follows. In Section~\ref{sec.prob-formulation}, we formulate the riblet shape-optimization problem as a constrained $\cH_2$ state-feedback synthesis problem based on the linearized NS model and volume-penalization framework, and summarize the associated structural and physical constraints. In Section~\ref{sec.Approach.sigdis}, we introduce the signed-distance function parameterization and reformulate the optimization problem in terms of parameterization weights. In Section~\ref{sec.alg}, we develop a customized method of multipliers (MM) algorithm for solving the reformulated problem. In Section~\ref{sec.Results}, we construct the library of riblet geometries used in the morphing procedure and analyze solutions obtained from both deterministic and randomized initializations. Finally, Section~\ref{sec.conclusions} summarizes the main findings and outlines future directions.


\section{Problem formulation}
\label{sec.prob-formulation}

\begin{figure}
        \begin{tabular}{rcrc}
        \hspace{-.2cm}
        \subfigure[]{\label{fig.channelriblets}}
        &&
        \hspace{-.2cm}
        \subfigure[]{\label{fig.ribshapes}}
        &
        \\[-.3cm]
	&
        \hspace{-.5cm}
        \includegraphics[width=0.26\textwidth]{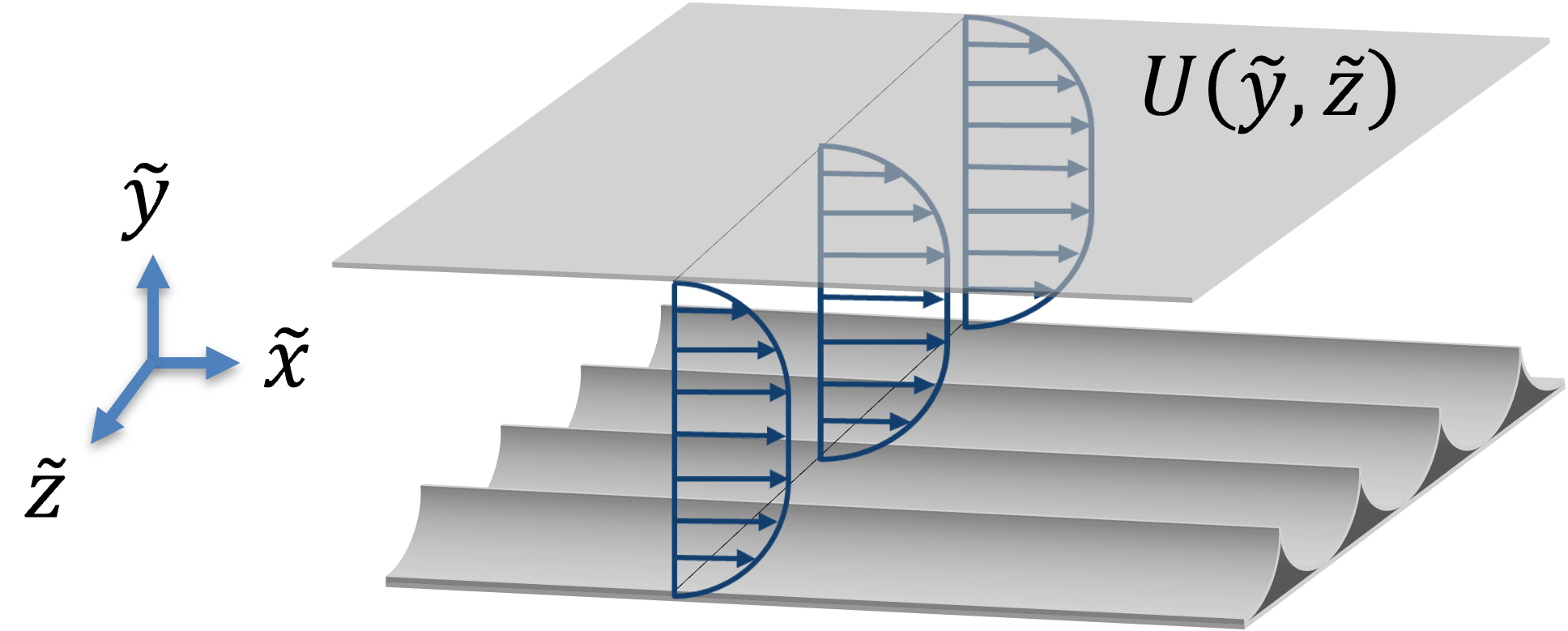}
        &&
        \hspace{-.5cm}
        \includegraphics[width=0.18\textwidth]{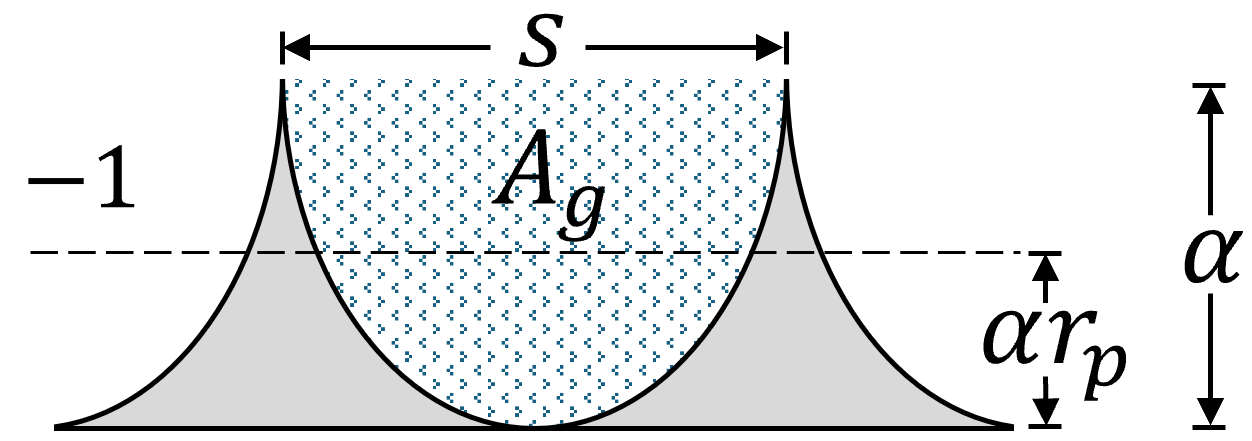}
        \end{tabular}
        \vspace{-.2cm}
        \caption{(a) Channel flow with riblets on the lower wall together with the mean velocity in blue. (b) Semi-circular riblets of spanwise frequency $\omega_z$, height $\alpha$, groove cross-sectional area $A_g$, and spacing $s=2\pi/\omega_z$. The parameter $r_p$ is the proportion of the riblet height below $\tilde{y}=-1$.}
        \vspace{-.4cm}
\end{figure}

The dynamics of a pressure-driven turbulent channel flow over a riblet-mounted surface (Fig.~\ref{fig.channelriblets}) are given by the NS equations,
\begin{subequations}
\begin{align}
\label{eq.NSeqn}
\ba{rcl}
    \bu_t
    &\!\!=\!\!&
    -
    \left( \bu \cdot \nabla \right) \bu
    \,-\,
    \nabla P
     \,+\,
    \dfrac{1}{Re_\tau}\, \Delta \bu
    \\
    0
    &\!\!=\!\!&
    \nabla \cdot \bu
\ea
\end{align}
subject to no-slip/no-penetration boundary conditions,
\begin{align}
    \label{eq.BCs}
    \!\!\bu(x,y=1,z,t)\,=\, 0,~\; \bu(x,y=-1+r(z),z,t)\,=\, 0\!\!
\end{align}
\label{eq.NSeqn-BCs}
\end{subequations}
that capture the effect of the streamwise-constant corrugation of the lower wall through the shape function $r(z)$. Here, $\bu$ is the three-dimensional velocity vector, $P$ is pressure, $\nabla$ is the gradient operator, $\Delta = \nabla\cdot \nabla$ is the Laplacian, $t$ is time, and the Reynolds number $Re_\tau \DefinedAs u_\tau h/\nu$ results from non-dimensionalization of length by the channel half-height $h$, velocity by the friction velocity $u_\tau = \sqrt{\tau_w/\rho}$, time by $h/u_\tau$, and pressure by $\rho u_\tau^2$. Here, $\tau_w$ is the wall-shear stress (averaged over horizontal directions and time), $\rho$ is the fluid density, and $\nu$ is the kinematic viscosity.

The effect of engineered surface corrugations (e.g., riblets) is typically quantified via their impact on skin-friction drag, which is governed by the near-wall gradient of the turbulent mean velocity, and by the energy contained in turbulent fluctuations. Here, we focus on the latter and optimize the surface geometry $r(z)$ to reduce the energy of the turbulent velocity fluctuations. The dynamics of zero-mean velocity and pressure fluctuations, $\bv=[\,u\,~v\,~w\,]^T$ and $p$, about the turbulent mean $\bar{\bu}$, which completes the decomposition of the flow field in Eq.~\eqref{eq.NSeqn}, i.e., $\bu = \bar{\bu} + \bv$ and $P = \bar{P} + p$, are described by the linearized NS equations, 
\begin{align}
	\non
    		\bv_t
    		&\; = \;
    		-
    		\left( \bar{\bu}\cdot \nabla  \right) \bv
     		\; - \;
    		\left( \bv \cdot \nabla \right) \bar{\bu}
    		\; - \;
    		\nabla p \;-\; {K}^{-1}\bv
            \\
            \non
    		&\hspace{1cm} +\,
    		\dfrac{1}{Re_\tau} \nabla\cdot\left((1\,+\,\nu_T)( \nabla\bv \;+\; (\nabla\bv)^T )\right) \;+\; \mathbf{f},
    		\\
		\label{eq.fluc-dyn}
    		0
		&\; = \;
    		\nabla \cdot \bv
\end{align}
In these equations, we follow~\cite{ranzarjovJFM21} in accounting for riblets not by enforcing boundary conditions through mesh deformation, but by penalizing the fluctuation dynamics using a permeability field $K(y,\,z)$ that conforms to the lower-wall geometry. This field damps fluctuations within the riblet grooves: $K \to 0$ inside the riblet structures, which forces $\bv\to 0$, while $K \to \infty$ in the fluid region, recovering the unmodified equations. Due to this, we can now impose standard homogeneous Dirichlet and Neumann boundary conditions on $\bv$ at both the upper and lower walls. In Eq.~\eqref{eq.fluc-dyn}, $\mathbf{f}(t)$ is a zero-mean stochastic forcing that is added to the momentum equations to excite a stochastic response from the linearized dynamics. We model this source of excitation as a white-in-time process with covariance $M \succ 0$, i.e., $\mathbf{E} \left(\mathbf{f}(t) \otimes \mathbf{f}(\tau)\right) = M\delta(t-\tau)$, where $\delta$ denotes the Dirac delta distribution and $\otimes$ the tensor (outer) product. Finally, the mean velocity $\bar{\bu}$ is itself obtained as the solution to the steady-state equations,
\begin{align}
	\non
	0
    &\; = \;
    -\,
    \left( \bar{\bu} \cdot \nabla \right) \bar{\bu}
    \;-\;
    \nabla \bar{P} \;-\; {K}^{-1}\,\bar{\bu}
    \\
    \non
    & \hspace{1cm} + \,
    \dfrac{1}{Re_\tau} \nabla\cdot\left((1\,+\,\nu_T)( \nabla\bar{\bu} \;+\; (\nabla\bar{\bu})^T )\right),
    \\
    \label{eq.meanflow}
    0
    &\; = \;
    \nabla \cdot \bar{\bu}
\end{align}
which result from application of the turbulent viscosity hypothesis~\cite{mcc91} to the Reynolds-averaged NS equations. Here, $\nu_T(y)$ is the turbulent eddy viscosity normalized by molecular viscosity; see~\cite{ranzarjovJFM21} for additional details. For the small-size riblets considered in this paper, we consider the flow in the grooved region of the riblet structures (Fig.~\ref{fig.ribshapes}), which is dominated by viscosity, to be laminar, i.e., $\nu_T=0$ for $y \leq -1$ in both Eqs.~\eqref{eq.fluc-dyn} and~\eqref{eq.meanflow}.

For streamwise-constant spanwise-periodic surface corrugation, $K^{-1}(y,z)$ can be expanded using the Fourier transform,
\begin{align}
\label{eq.invK}
	K^{-1}(y,z)
	\;=\;
	\sum_{\mathrm{m}\in\bbZ}\,a_\mathrm{m}(y) e^{({\mri \mathrm{m}\,\omega_z z})}
\end{align}
Here, $\omega_z$ is the spatial frequency of the surface corrugation and the Fourier coefficients $a_\mathrm{m}(y)$ (functions of $y$) specify their shape and height in the vertical direction. Based on this, the solution to Eqs.~\eqref{eq.meanflow} adopts a similar harmonic form, i.e.,
\begin{align}
\label{eq.baseflow}
	\bar{\bu}(y,\,z)
	\;=\;
	\sum_{\mathrm{m}\in\bbZ}\,\bar{\bu}_\mathrm{m}(y) \exp({\mri\mathrm{m}\,\omega_z z}).
\end{align}

\begin{figure*}
\begin{align}
\label{eq.BC-smoothness}
\ba{rclrcl}
\hat{v}_n(k_x,y=-1^+,\theta,t)
&\!\!=\!\!&
\hat{v}_n(k_x,y=-1^-,\theta,t), \quad 
&\dfrac{\partial \hat{v}_n}{\partial y}(k_x,y=-1^+,\theta,t)
&\!\!=\!\!&
\dfrac{\partial \hat{v}_n}{\partial y}(k_x,y=-1^-,\theta,t)
\\[.25cm]
\dfrac{\partial^2 \hat{v}_n}{\partial y^2}(k_x,y=-1^+,\theta,t)
&\!\!=\!\!&
\dfrac{\partial^2 \hat{v}_n}{\partial y^2}(k_x,y=-1^-,\theta,t), 
&\dfrac{\partial^3 \hat{v}_n}{\partial y^3}(k_x,y=-1^+,\theta,t) 
&\!\!=\!\!&
\dfrac{\partial^3 \hat{v}_n}{\partial y^3}(k_x,y=-1^-,\theta,t)
\\[.25cm]
\hat{\eta}_n(k_x,y=-1^+,\theta,t)
&\!\!=\!\!& 
\hat{\eta}_n(k_x,y=-1^-,\theta,t), \quad 
&\dfrac{\partial \hat{\eta}_n}{\partial y}(k_x,y=-1^+,\theta,t)
&\!\!=\!\!&
\dfrac{\partial \hat{\eta}_n}{\partial y}(k_x,y=-1^-,\theta,t)
\ea
\end{align}
\hrulefill
\end{figure*}

For small riblets, the solution to the nonlinear equations~\eqref{eq.meanflow} obtained via Newton's method has a single streamwise component, $\bar{\bu} = [\,U(y,z) \,~0\,~0\,]^T$, which significantly simplifies the 
equations by bringing them into a linear form.

The solution to Eqs. (2) shares the harmonic structure of $K^{-1}$ and $\bar{\bu}$; while the operators in Eqs.~\eqref{eq.fluc-dyn} are invariant in $x$, they are spatially periodic in $z$. By adopting Bloch wave theory~\cite{odekel64,benliopap78} to parameterize the spatially periodic dynamics in the Fourier domain and applying a standard pressure-elimination conversion~\cite{schhen01}, the PDEs~\eqref{eq.fluc-dyn} can be recast in evolution form as
\begin{align}
\label{eq.lnse}
    \bpsi_t (\bk,y,t)
    \;=\;
    \bA(\bk)\, \bpsi(\bk,y,t)
    \, + \,
    \bB(\bk)\, {\bf f} (\bk,y,t)
\end{align}
where the state $\bpsi$ consists of the wall-normal velocity $v$ and vorticity $\eta \,=\, \partial_z u \,-\, \partial_x w$. Here, $\bpsi$, and $\mathbf{f}$ are bi-infinite column vectors parameterized by the wavenumber pair $\bk = (k_x,\theta)$, with $k_x\in \mathbb{R}$ and $\theta \in [ 0,\omega_z)$ denoting the streamwise wavenumber and spanwise wavenumber offset, e.g., for every $\bk$, $\bpsi(\bk,y,t)=\col\left \{ \hat{\bpsi}_n(k_x,y,\theta,t) \right \}_{n\in\mathbb{Z}}$ with $\hat{\bpsi}_n = [\,\hat{v}_n \,~ \hat{\eta}_n\,]^T$ for any integer $n$. As a result of this parameterization, the spanwise wavenumber is given by $\theta_n=\theta + n\,\omega_z$~\cite{ranzarjovJFM21}. At both the top and the bottom boundaries, homogeneous Dirichlet and Neumann conditions are imposed on $\hat{v}_n$, and homogeneous Dirichlet conditions are imposed on $\hat{\eta}_n$. The adopted volume penalization method should automatically enforce immersed boundary conditions on the non-smooth lower wall without the need for additional boundary conditions. However, to avoid numerical oscillations in practice, the computational domain is divided into two parts using block operators~\cite{aurtre17} with smoothness enforced at the interface; see Eqs.~\eqref{eq.BC-smoothness}. 

In Eq.~\eqref{eq.lnse}, the system matrices, $\bA(\bk)$ and $\bB(\bk)$,  are bi-infinite with operator-valued elements in $y$~\cite[Appendix A]{ranzarjovJFM21}. We employ a change of variables to obtain a state-space representation in which the kinetic energy is determined by the Euclidean norm of the state vector in a finite-dimensional approximation of the evolution model~\cite[Appendix A]{zarjovgeoJFM17}. A pseudospectral scheme is used for discretizing the wall-normal operators in both the mean flow equations~\eqref{eq.NSeqn} and the linearized dynamics~\eqref{eq.lnse}~\cite{weired00}. This yields the finite-dimensional approximation,
\begin{align}
\label{eq.lnse-finite}
    \bpsi_t (\bk,t)
    \;=\;
    A(\bk)\, \bpsi(\bk,t)
    \, + \,
     B(\bk)\, {\bf f} (\bk,t)
\end{align}
where $\bpsi$ and ${\bf f}$ are now vectors whose dimensions are determined by the adopted wall-normal discretization; if $N_l$ and $N_u$ collocation points are used to discretize the lower and upper wall-normal regimes, respectively, and $N_h$ harmonics are used in the Fourier Series expansion~\eqref{eq.invK}, $\bpsi(\bk, t)$ is a vector of $2\times N_h \times (N_l + N_u)$ complex-values entries. Moreover, matrices 
$A(\bk)\in\bbC^{2 N_h N_y\times 2N_h N_y}$ and $B(\bk)\in\bbC^{2N_h N_y\times 3N_h N_y}$, with $N_y = N_l + N_u$, 
are discretized versions of the operators in Eq.~\eqref{eq.lnse} after incorporating the aforementioned change of coordinates.
The state matrix $A(\bk)$ can be decomposed into a nominal component associated with the dynamics of a channel flow with smooth walls ($A_\mathrm{s}$) and a component that represents the riblet-induced deviation from the smooth-wall channel flow dynamics ($\Gamma_{a_\mathrm{m}}$), i.e., $A = A_\mathrm{s} + \Gamma_{a_\mathrm{m}}$. Here, $\Gamma_{a_\mathrm{m}}$ is a structured matrix in which the Fourier coefficients of $K^{-1}$ and $\bar{\bu}$ appear as diagonals of submatrices; see Appendix~\ref{subsec:strcgamfeedmat} for the description of $\Gamma_{a_\mathrm{m}}$.  

The permeability function $K(y,z)$, whose inverse appears as a coefficient both directly in the fluctuation dynamics~\eqref{eq.fluc-dyn} and indirectly through the mean velocity $\bar{\bu}$ (cf.~Eq.~\eqref{eq.meanflow}), can be tailored to capture the structure of arbitrarily shaped riblets defined by roughness functions $r(z)$~\cite{ranzarjovJFM21}. On the other hand, the appearance of $K^{-1}(y,z)$ as a multiplicative gain in the dynamics of velocity fluctuations (i.e., $- {K}^{-1}\bv$ term in Eqs.~\eqref{eq.fluc-dyn}) highlights an opportunity to design {\em optimally} rough surfaces for suppressing energy via optimal control theory. 

\subsection{Constrained optimal control problem}

We examine the wall-normal structure of $K^{-1}(y,z)$ for a streamwise-constant, spanwise-periodic roughness pattern of prescribed height and spanwise frequency $\omega_z$. 
Our goal is to determine the feedback gain matrix $\Gamma_{a_\mathrm{m}}$ that minimizes the quadratic cost functional 
\begin{align}
\label{eq.cost}
    J(\Gamma_{a_\mathrm{m}}) \DefinedAs
    \begin{cases} 
      \ds{\int\trace (Q\,X \,+\, \Gamma_{a_\mathrm{m}}^* R\, \Gamma_{a_\mathrm{m}} X )\mrd \bk}, & \Gamma_{a_\mathrm{m}}\in\, {\cal C}\\
      \infty, & \text{otherwise}
   \end{cases}
\end{align}
where $X(\bk) \DefinedAs \mathbf{E} \left(\bpsi(\bk,\cdot,t) \otimes \bpsi(\bk,\cdot,\tau)\right)$ is the state autocorrelation matrix,
$Q=Q^* \succ 0$ and $R=R^* \succ 0$ specify penalties on the state and riblet-induced dynamical modification, respectively, ${\cal C}$ denotes the set of structured stabilizing feedback matrices that satisfy all physical requirements levied on the system (Sec.~\ref{sec.optimization-constraints}), and $*$ is the complex-conjugate transpose. Both the stability criterion and consistency with an admissible state autocorrelation $X \succ 0$ reduce to an algebraic constraint on $\Gamma_{a_\mathrm{m}}$ and $X$, namely
\begin{align}
\label{eq.lyap}
    \left(A_\mathrm{s} + \Gamma_{a_\mathrm{m}}\right) X
    \;+\;
    X \left(A_\mathrm{s} + {\Gamma_{a_\mathrm{m}}}\right)^*
    \;+\; B\, M B^* \;=\; 0.
\end{align}

In a standard unstructured state-feedback $\mathcal{H}_2$ optimal control problem, the optimal gain matrix $\Gamma_{a_\mathrm{m}}$ can be obtained by solving a single algebraic Riccati equation (equivalently, the LQR ARE); see, e.g.,~\cite[Sec.~9.2, Eqs.~(9.12)--(9.13)]{skopos07}. However, as described in the following subsection, the structural constraints 
imposed on our problem render it nonconvex, precluding solution by standard Riccati-based methods.

\subsection{Optimization constraints}
\label{sec.optimization-constraints}

There are a variety of physical constraints that comprise the feasibility set $\mathcal{C}$. We classify these constraints into three categories: Harmonic-consistency constraints; constant bulk and equal spacing; consistent penalization; and structural connectivity.

\begin{figure}
    \centering
    \includegraphics[width=0.65\linewidth]{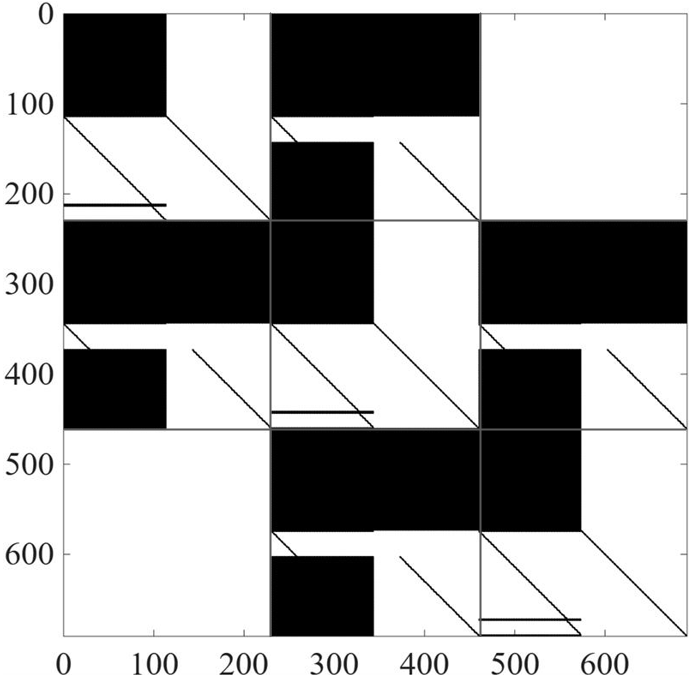}
    \caption{Sparsity pattern of entries in the feedback gain matrix $\Gamma_{a_\mathrm{m}}$ in the riblet dynamical augmentation. Gray lines separate submatrices corresponding to various harmonic interactions when $N_y=121$ and $N_h=3$.}
    \label{ribletStructure}
\end{figure}

\vsp
\subsubsection{Harmonic-consistency constraints}
As noted above, the Fourier coefficients  $a_\mathrm{m}(y)$ and $\bar{\bu}_\mathrm{m}(y)$ appear on the diagonal of the submatrices that comprise $\Gamma_{a_\mathrm{m}}$.  Figure~\ref{ribletStructure} illustrates the sparsity pattern of the feedback matrix $\Gamma_{a_\mathrm{m}}$ if the harmonic expansions~\eqref{eq.invK} and~\eqref{eq.baseflow} are truncated at 3 harmonics. Each sub-matrix encodes the dynamical coupling between (i) harmonics of the state fluctuations $\hat{\bpsi}_n$, and (ii) harmonics of the penalization term $K^{-1}(x,z)$ and the base flow $\bar{\bu}(y,z)$, which are coefficients in Eqs.~\eqref{eq.fluc-dyn}; see~\cite{ranzarjovJFM21} for details. Consequently, any optimization-based construction of $\Gamma_{a_\mathrm{m}}$ must be posed at the level of the Fourier profiles of the penalization function $a_\mathrm{m}(y)$. This is necessary not only to preserve the intended harmonic interactions, but also to enforce consistency with the base-flow harmonics that satisfy
\begin{align}
\label{meanRef}
    \ba{l}
    \hspace{-.4cm}
    \left[(1+\nu_T)\left(\partial_y^2-\mathrm{m}^2 \omega_z^2\right) \,+\, \nu_T'\,\partial_y \,-\,a_0\right]\hat{U}_\mathrm{m} 
    \\[.15cm]
    \hspace{1.6cm}
    ~+\; \ds{\sum_{\mathrm{n} \in \bbZ}}\, a_\mathrm{n} \, \hat{U}_\mathrm{m-n} 
    \;=\; 
    \begin{cases} 
      Re_{\tau}\bar{P_x} & \mathrm{m} = 0 \\
      0 & \mathrm{m}\neq 0 
    \end{cases}
    \ea
\end{align}

Here, $\bar{U}(y)\DefinedAs\col\left \{ \hat{U}_\mathrm{m}(y) \right \}_{\mathrm{m}\in\mathbb{Z}}$ is a bi-infinite vector of the streamwise base-flow harmonics. Since these harmonics appear in the structured feedback matrix $\Gamma_{a_\mathrm{m}}$, a consistency constraint must be enforced to ensure that Eq.~\eqref{meanRef} holds whilst adhering to the structure shown in Fig.~\ref{ribletStructure}. 

\vsp
\subsubsection{Constant bulk and equal spacing}
\label{sec.constant-bulk}

To obtain a tractable design process with reduced degrees of freedom, we assume that the riblet spacing is fixed, i.e., $\omega_z$ is fixed. Nevertheless, an additional constraint is necessary to prevent the optimization from reducing kinetic energy simply by decreasing the mass flux, which would inadvertently increase surface drag. To this end, we enforce the constant bulk-flow constraint,
\begin{align}
    0 \;=\; U_B \,-\, \dfrac{1}{1-P_x}\int_{0}^{\frac{2\pi}{\omega_z}} \!\! \int_{-1-\alpha r_p}^{1}\! U(y,z)\,\mrd y\, \mrd z,
\label{eq.bulkIntegral}
\end{align}
where $U_B$ is the bulk flux in a turbulent channel flow with smooth walls. Here, $\alpha$ denotes the riblet height, while $r_p$ represents the ratio of the riblet height to the thickness of the laminar subregion; see Fig.~\ref{fig.ribshapes}.

\vsp
\subsubsection{Consistent penalization}
\label{sec.consistent-penalization}
A property of the permeability function $K^{-1}(y,z)$ that is not inherently encoded in the Fourier coefficients $a_\mathrm{m}(y)$ is binary masking. Specifically, this property requires $K^{-1}(y,z)$ to assume only one of two values at any location in the $y–z$ plane, i.e.,
\begin{align}
    K^{-1}(y,z) 
    \;=\; 
    \begin{cases} 
      R_0 & \text{if $(y,z)$ is inside the riblet}\\
      0 & \text{if $(y,z)$ is outside the riblet}
   \end{cases}
   \label{Consistent_Pen}
\end{align} 
Therefore, throughout the optimization process, the Fourier profiles $a_\mathrm{m}(y)$ must belong to a class of functions that preserve this property. 

If this penalization constraint is satisfied, the riblet structure will be physically realizable as it will have a well-defined edge. Equation$~\eqref{Consistent_Pen}$ describes an instantaneous jump at the edge of the riblet. In practice, smooth transitions are necessary to maintain physical consistency with wall-bounded turbulent flows. If there is too sharp of a transition, the feedback configuration $A+\Gamma_{a_\mathrm{m}}$ may become unstable. On the other hand, in the case that the edge of the riblet is too smoothed out, spurious velocity fluctuation values can appear within the riblet structures. This limitation of immersed boundary methods can be seen by describing the gradient at the edge of the roughness structure using as a hyperbolic tangent function~\cite[Eq.~(2.9)]{ranzarjovJFM21}. 

\vsp
\subsubsection{Structural connectivity}
\label{sec.connectivity}

\begin{figure}
\vspace{.6cm}
\begin{tabular}{rcc}
	\hspace{-.4cm}
	\begin{tabular}{c}
	\vspace{.4cm}
	\rotatebox{90}{\footnotesize $y$}
	\end{tabular}
&
	\hspace{-0.9cm}
	\begin{tabular}{c}
            \includegraphics[width=.22\textwidth]{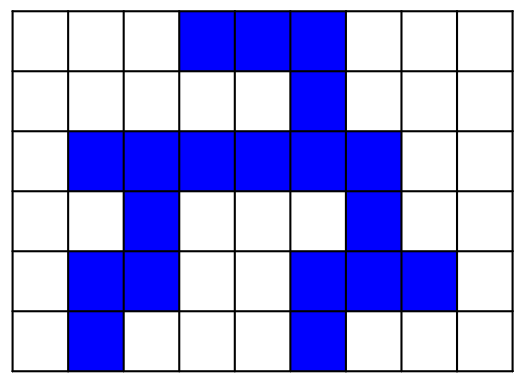}
		\\[-.1cm]
		{\small $z$}
	\end{tabular}
&
	\hspace{-.6cm}
	\begin{tabular}{c}
            \includegraphics[width=.22\textwidth]{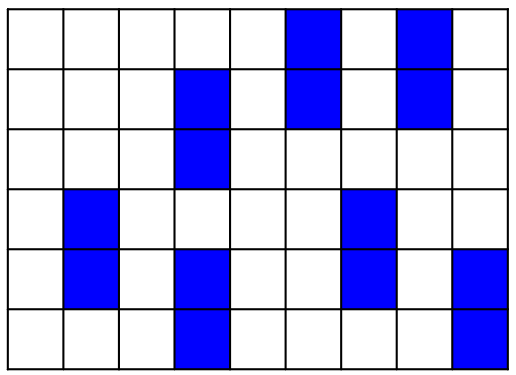}
		\\[-.1cm]
		{\small $z$}
	\end{tabular}
	\\[-.2cm]
&
\hspace{-0.9cm}
	\subfigure[]{\label{fig.connected}}
&
\hspace{-0.6cm}
	\subfigure[]{\label{fig.disconnected}}
\end{tabular}
\vspace{-.4cm}
\caption{Sparsity pattern of resistance function $K^{-1}(y,z)$ for (a) a fully connected manufacturable design; and (b) a vertically disconnected design.}
\end{figure}

Finally, we must maintain vertical support (connectivity) to the wall: every point within the riblet must have a fully enclosed path back to the wall. Consider a finely discretized domain, refined to grid convergence. Because each cell in the discretization is permitted to take values of either 0 or $R_0$, isolated (disconnected) solid cells can arise. The left panel of Fig.~\ref{fig.connected} illustrates a hollow yet fully connected riblet: every point within the solid phase admits a continuous, enclosed path to the wall. By contrast, the right panel shows a disconnected configuration, which is not manufacturable. The admissibility of local cell patterns depends on proximity to riblet edges, implying that connectivity cannot be enforced purely through local rules; it is inherently a global constraint.

To impose this requirement, prior work in additive manufacturing has introduced pointwise topological constraints. One common approach injects support structures for unsupported regions: when a “floating island” appears, the constraint violation triggers the creation of a support beam~\cite{gaygue16}. In our setting, however, enforcing such a pointwise condition in Fourier space would introduce a large number of nonlinear constraints (one per grid point) and may also generate extraneous structures that are irrelevant outside the additive-manufacturing context.
As an alternative, Li et al.~\cite{li16} proposed a virtual-temperature formulation that guarantees connectivity by solving an auxiliary Laplace-type problem at each iteration. While it is possible to impose this virtual temperature constraint within an optimization loop, doing so introduces numerous embedded constraints; translating each of these into a corresponding constraint on the feedback matrix $\Gamma_{a_\mathrm{m}}$ can be cumbersome. 

We instead adopt parameter-gradient methods applied to the $\cH_2$ cost, which have primarily been employed outside of structured control problems \cite{pet13,gomegomonmic18}. Within this framework, we pursue a shape-optimization approach that preserves wall connectivity by construction. In particular, the design-by-morphing method introduced in the next section uses a change of variables that maps the baseline optimization field $a_\mathrm{m}(y)$ to a lower-dimensional set of morphing parameters $w$. Whether enforcing connectivity through a topologically constrained, grid-based formulation or via shape optimization—the central modeling decision is identical: how the riblet boundary (edge) is defined and parameterized.

\section{Approach: signed-distance function tuning}
\label{sec.Approach.sigdis}
We enforce uniform penalization (Sec.~\ref{sec.consistent-penalization}) and compactness (Sec.~\ref{sec.connectivity}) by employing a variable transformation that brings the problem into an equivalent shape optimization form. This approach produces a morphed riblet geometry using a linear combination of signed-distance functions. To solve the resulting optimization problem, we develop a customized algorithm, which  uses a parameter-gradient method to minimize the $\mathcal{H}_2$ norm of the feedback synthesis. 

\begin{figure}
    \centering
    \includegraphics[width=0.99\linewidth]{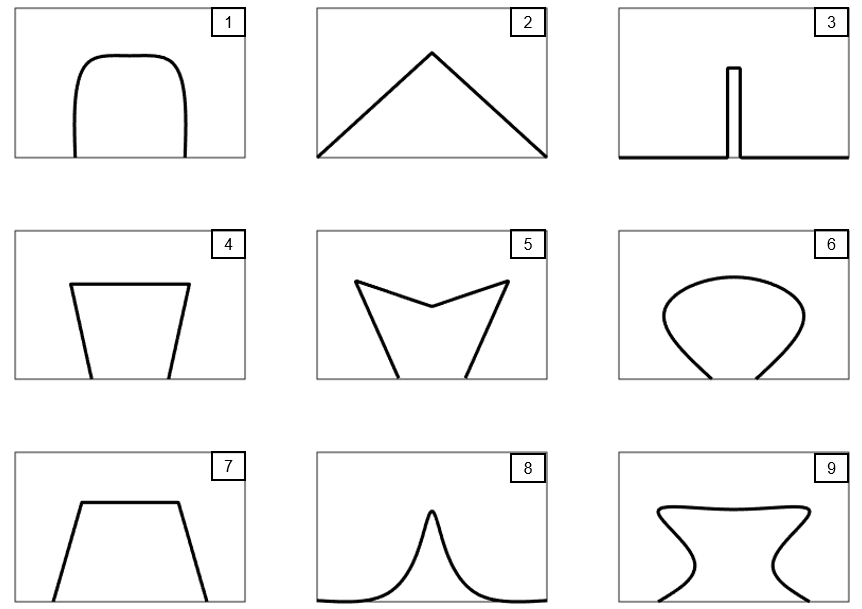}
    \caption{Visualization of the library of nine riblet shapes over which the optimization is performed.}
    \label{parametricCurves}
\end{figure}

\subsection{Surface representation using signed-distance functions}
Design-by-morphing has recently been combined with large-eddy simulations and Bayesian optimization to identify riblet geometries that are optimal for drag reduction~\cite{leemoalimgumar24}. 
The simulation-free method we develop in this section adapts existing tools in design by morphing to $\mathcal{H}_2$ optimal control under physical constraints. In this method, a morphed riblet shape is achieved by parameterizing the riblet geometry over a library of curves (Fig.~\ref{parametricCurves}). In order to find an optimal morphed shape from this library, one might first consider applying a standard linear combination to the library of resistance functions corresponding to these curves, i.e., 
\begin{align}
\label{eq.weightedK}
    K^{-1}(y,z) \;=\; \sum_i w_i\, K^{-1}_i(y,z).
\end{align}
 However, using this method to combine the basis shapes violates the consistent penalty constraint. This is because the intersection of any two resistance functions would contain twice the penalty. To resolve this issue, we use signed-distance functions, which were first developed in the computer graphics community for morphing binary masks through linear combinations~\cite{malsetvem02}.

\begin{figure}
\begin{tabular}{rcc}
	\hspace{-.6cm}
	\begin{tabular}{c}
	\vspace{.4cm}
	\rotatebox{90}{\footnotesize $y$}
	\end{tabular}
&
	\hspace{-0.8cm}
	\begin{tabular}{c}
            \includegraphics[width=.2175\textwidth]{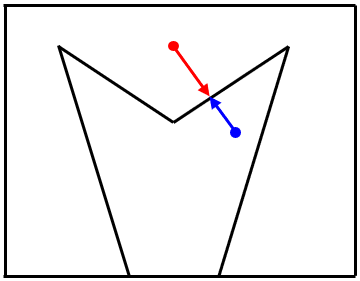}
		\\[-.1cm]
		{\small $z$}
	\end{tabular}
&
	\hspace{-.7cm}
	\begin{tabular}{c}
            \includegraphics[width=.23\textwidth]{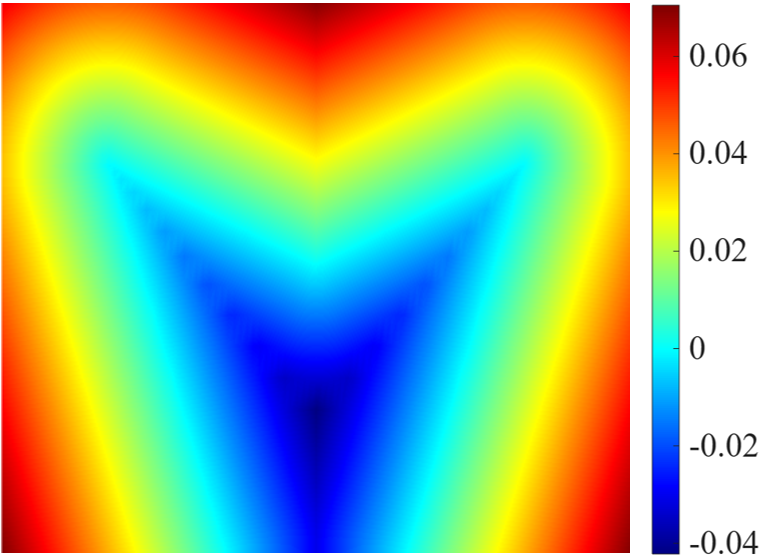}
		\\[-.1cm]
		{\hspace{-1cm} \small $z$}
	\end{tabular}
	\\[-.2cm]
&
	\hspace{-1.25cm}
	\subfigure[]{\label{fig.ParametricIllustration}}
&
	\hspace{-1.75cm}
	\subfigure[]{\label{fig.SDFIllustration}}
\end{tabular}
\vspace{-.4cm}
    \caption{Illustration of (a) the 2D shape function of a riblet structure; and (b) the corresponding signed-distance function.}
\end{figure}

Any parameterized curve can be converted into a two-dimensional signed-distance function that describes its boundary. For any point $(y,z)$, the signed-distance function $\phi(y,z)$ gives the distance to the nearest point on a shape's boundary: it is negative inside the shape and positive outside. Figures~\ref{fig.ParametricIllustration} and ~\ref{fig.SDFIllustration} illustrates this construction for a representative curve. 
Using this representation, the Fourier coefficients $a_\mathrm{m}$ in Eq.~\eqref{eq.invK} can be computed from the Fourier transform of a sigmoid applied to a linear combination of signed-distance functions $\phi_i$, i.e.,
\begin{align}
\label{eq.SDFVariableChange}
    a_\mathrm{m}(y)\;=\;{\cal FT}\left\{\sigma\left(\sum_iw_i\,\phi_i(y,z)\right)\right\}.
\end{align}
Here, $\w_i \in [0,1]$ are tuning weights, $\sigma(x) \DefinedAs 1/(1 + \mre^{x / \eps}) $ is a sigmoid function with $\eps>0$ controlling the transition rate between fluid and solid interfaces, and ${\cal FT\{\cdot\}}$ denotes the discrete Fourier transform in the spanwise direction. If all weights $w_i$ sum to one, the convex combination $\sum_i w_i \phi_i$ yields riblet shapes obtained by interpolating among curves in the library. Note that the wall-normal discretization of the signed-distance functions $\phi_i$ is chosen to match that of the linearized flow model. 

\subsection{Problem reformulation}
\label{sec.prob-form-Jw}
The expression of optimization variables $a_\mathrm{m}$ provided by the weighted sum of signed-distance functions (Eq.~\eqref{eq.SDFVariableChange}) transforms the constrained optimal control problem described in Sec.~\ref{sec.prob-formulation} into the following form
\begin{align}
\label{eq.problem-w}
\hspace{-.1cm}
\ba{cl}
	\minimize & J(\bw)
	\\[-.2cm]
	\bw
	\\
	\subject 
    &
    0 \;=\; \ds{\int_{0}^{\frac{2\pi}{\omega_z}} \!\!\! \int_{-1-\alpha r_p}^{1}\! U_\bw(y,z)\,\mrd y\, \mrd z}
    \\[.45cm]
    &
	0 \;=\; 1 \,-\, \mathds{1}^T\bw
    \\[.1cm]
    & 
    0\,\leq\, \bw \,\leq\, 1
\ea
\end{align}
where the optimization variable $\bw = \mathrm{vec}\{w_i\}$ denotes the vector of tuning weights $w_i$, $U_\bw(y,z)$ denotes riblet-induced changes to the streamwise component of the base flow $\bar{\bu}$, $\mathds{1}$ is the vector of ones of the same length as $\bw$ and
the dependence of the quadratic cost function $J$ on weights $w_i$ follows from its dependence on $\Gamma_{a_\mathrm{m}}$ (Eq.~\eqref{eq.cost}) and Eq.~\eqref{eq.SDFVariableChange}; see Appendix~\ref{subsec:strcgamfeedmat} for details.
In this formulation, all constraints highlighted in Sec.~\ref{sec.optimization-constraints} are implicitly satisfied but the constant bulk constraint (Sec.~\ref{sec.constant-bulk}). The linear equality constraint on $\bw$ ensures a convex combination and the constant bulk flux constraint is nonlinear. The problem is nonconvex due to the nonconvexity of the sigmoid function (Eq.~\eqref{eq.SDFVariableChange}) through which the weights $\bw$ are passed onto the Fourier coefficients in $\Gamma_{a_\mathrm{m}}$ and nonlinearity of the constraints on the bulk flux. As a result, the optimization landscape does not necessarily have a global minimizer. This reflects the dependence of solutions on initial conditions; see Sec.~\ref{sec.randomseed}. We next develop a customized optimization algorithm to solve problem~\eqref{eq.problem-w}.

\section{Optimization algorithm}
\label{sec.alg}

We employ the MM algorithm~\cite[Algorithm 17.4]{nocwri06} to solve the constrained optimization algorithm outlined in the previous section. This algorithm is a dual-ascent optimization scheme which has wide application in solving constrained nonlinear programming problems~\cite{ber82,ber99,nocwri06,zarmohdhigeojovTAC20}. The MM algorithm for solving problem~\eqref{eq.problem-w} is given by
 \begin{subequations}
 \begin{align}
 \label{w.subprob}
    \bw^{k+1} \;\DefinedAs\; \argmin_{0\leq \bw\leq 1}\; \mathcal{L}_{\rho^k}(\bw ;\lambda^k_1, \lambda^k_2) 
 \\[.15cm]
 \label{lambda.subprob}
    \mathbf{\lambda}^{k+1}_i \;\DefinedAs\; \lambda^k_i \,+\, \rho^k_i\, c_i 
 \end{align}
 \end{subequations}
 where $\cL_{\rho^k}$ is the augmented Lagrangian associated with linear constraints at the $k$th iteration, i.e., 
\begin{align}
\label{eq.augLagrangian}
    \mathcal{L}_{\rho}(\bw;\lambda_1,\lambda_2)\;\DefinedAs\; J(\bw) \,+\, 
    \sum_{i}\mathbf{\lambda}_{i}\,
	c_i
    \,+\, 
    \dfrac{1}{2} \sum_{i}\mathbf{\rho}_{i}\,
	c_i^2.
\end{align}
Here, $\lambda_i \in \bbR$ ($i=1,2$) are Lagrange multipliers, $\rho_i$ ($i=1,2$) are positive scalars, $c_i$ are entries of the vector of equality constraints,
\begin{align*}
    \mathbf{c} 
    \;\DefinedAs\;
    \begin{bmatrix}
    \ds{\int_{0}^{\tfrac{2\pi}{\omega_z}} \!\!\!\! \int_{-1-\alpha r_p}^{1}\! U_\bw(y,z)\,\mrd y\, \mrd z}
    \\[.4cm]
    1\,-\, \mathds{1}^T\bw
    \end{bmatrix}.
\end{align*} 

The algorithm terminates when the primal $\Delta_p = \norm{\mathbf{c}}_2$ and dual $\Delta_d = \norm{\bw^{k+1}\,-\, \bw^k}_2$ residuals are small enough.

\subsection{Solution to the $\bw$-minimization problem~\eqref{w.subprob}}
For fixed $\{\lambda_1^k,\lambda_2^k,\rho_1^k,\rho_2^k\}$, minimizing the augmented Lagrangian with
respect to $\bw$ amounts to finding the minimizer of $\cL_{\rho^k}(\bw;\lambda_1^k,\lambda_2^k)$ subject to the box constraint $0\leq \bw\leq 1$. Since $\cL_{\rho^k}(\bw)$ is smooth, a projected gradient method can be used to find a minimizer of Eq.~\eqref{eq.augLagrangian} iteratively:
\begin{align*}
    \bw^{j+1} \;=\; {\cal P}_{[0,1]}\left(\bw^j \,-\, \alpha_{j}\nabla \cL_{\rho^k}(\bw^j)\right)
\end{align*}
where ${\cal P}_{[0,1]}(\cdot)$ denotes the projection of vector entries onto the feasible $[0,1]$ domain, $j$ is the inner gradient descent iteration counter, and $\alpha_j>0$ is the step size. The expression for the gradient of $J(\bw)$ is provided in Appendix~\ref{subsec:ParameterGradOpt}.

At each iteration $k$ of the projected gradient method, we determine the step size $\alpha_j$ via an adaptive Barzilai–Borwein (BB) initial
step size selection~\cite{golstubar14}, i.e.,
\begin{align}
	\label{eq.init-alpha}
	\alpha_{j,0}
	\;=\; 
	\left\{
	\ba{ll}
		\alpha_m~~
		& 
		\text{if}~\, \alpha_m/\alpha_s 
		\, > \, 
		1/2
		\\[.1cm]
		\alpha_s \,-\, \alpha_m/2 ~~
		&
		\text{otherwise}
	\ea 
	\right.
\end{align}
followed by backtracking to ensure sufficient descent, i.e., 
\begin{align*}
    \cL_{\rho^k}(\bw^{j+1}) 
    ~\leq~
    \cL_{\rho^k}(\bw^j) \,+\, \inner{\nabla \cL_{\rho^k}(\bw^j)}{\bw^{j+1} - \bw^{j}} 
    \\
    \,+\; \dfrac{1}{2\alpha_j} \norm{\bw^{j+1} - \bw^{j}}_2^2.
\end{align*}
In Eq.~\eqref{eq.init-alpha}, $\alpha_s$ and $\alpha_m$ are the ``steepest descent'' and ``minimum residual" step sizes, which are given by,
{\footnotesize \begin{align*}
	\ba{rcl}
	\alpha_s
	&\!\!\!=\!\!\!& 
	\dfrac{\inner{\bw^j - \bw^{j-1}}{\bw^j - \bw^{j-1}}}{\inner{\bw^j - \bw^{j-1}}{\nabla {\cL_{\rho^k}}(\bw^j) - \nabla {\cL_{\rho^k}}(\bw^{j-1})}}
	\\[.15cm]
	\alpha_m
	&\!\!\!=\!\!\!&
	\dfrac{\inner{\bw^j - \bw^{j-1}}{\nabla {\cL_{\rho^k}}(\bw^j) - \nabla {\cL_{\rho^k}}(\bw^{j-1})}}{\inner{\nabla {\cL_{\rho^k}}(\bw^j) - \nabla {\cL_{\rho^k}}(\bw^{j-1})}{\nabla {\cL_{\rho^k}}(\bw^j) - \nabla {\cL_{\rho^k}}(\bw^{j-1})}}.
	\ea
\end{align*}}
When $\alpha_s < 0$ or $\alpha_m < 0$, 
we use the following expression developed by Dai et. al. \cite{daialbyan15}: 
\begin{align*}
    \alpha^j
    \,=\,
    ||\bw^j-\bw^{j-1}||_2/||\nabla {\cL_{\rho^k}}(\bw^j)- \nabla {\cL_{\rho^k}}(\bw^{j-1})||_2
\end{align*}
For additional details, see~\cite[Sec. 4.1]{golstubar14}. Finally, we use a combined stopping criterion that terminates the projected gradient iterations when either the relative residual
\begin{align*}
	r_r^{j+1}
	= 
	\dfrac{||r^{j+1}||}{\max{\{||\nabla {\cL_{\rho^k}}(\bw^{j+1})||,||(\hat{\bw}^{j+1}-\bw^{j+1})/\alpha_j||\}} + \epsilon_r}
\end{align*}
or the normalized residual
\begin{align*}
	r_n^{j+1}
	\;=\;
	\dfrac{||r^{j+1}||}{||r^1|| + \epsilon_n}
\end{align*}
meet a desired tolerance. In the equations above, $\epsilon_r$ and $\epsilon_n$ are small positive constants and the residual is defined as
\begin{align*}
	r_n^{j+1}
	\;=\;
	\nabla \cL_{\rho^k}(\bw^{j+1}) \,+\, (\hat{\bw}^{j+1}-\bw^{j+1}) / \alpha_j
\end{align*}
where $\hat{\bw}^{j+1} \DefinedAs \bw^j-\alpha_j \nabla \cL_{\rho^k}(\bw^j)$. Note that the denominator of $r_r$ nearly vanishes when the gradient is small. In such cases, the first order optimality condition $(\nabla\cL_{\rho^k}(\bw^{j+1})=0)$ and $\norm{r_n}$ provide an appropriate stopping criteria; see \cite[Section 4.6]{parboy13} for additional information. 

\subsection{Lagrange multiplier and penalty parameter updates}
Customized MM for optimal riblet design is summarized in Algorithm~\ref{alg:cap}. The Lagrange multipliers $\lambda_i$ and penalty parameters $\rho_i$ are updated based on a slight variation of the procedure proposed in~\cite[Algorithm 17.4]{nocwri06}, which incorporates rules for the independent adjustment of the augmented Lagrangian penalty parameters~\cite{antkourid23,bassen23}. This procedure permits inexact solutions to subproblem~\eqref{w.subprob} and refined updates of the Lagrange multipliers and penalty parameters based on the adjustment of convergence tolerances on $\Delta_p$ and $\Delta_d$. 

\subsection{Computational complexity}
Each iteration of Algorithm~\ref{alg:cap} requires solving Lyapunov equations, performing matrix multiplications, evaluating traces, and carrying out other standard linear-algebraic operations. Since solving a single Lyapunov equation has worst-case computational complexity $O(n^3)$ and dominates the remaining operations, the overall computational complexity is $O(q_1q_2n^3)$, where $q_1$ denotes the number of outer-loop iterations and $q_2$ denotes the average number of inner-loop iterations required for convergence. 

\begin{algorithm}[h]
\caption{Customized MM algorithm}
\label{alg:cap}
\begin{algorithmic}
\STATE \textbf{input:} $A$, $B$, $V$, $Q$, $R$, and positive constants $\rho^0,\rho^1$, $\eta^0=(\rho^0)^{-0.1}$ $\delta_p^*$, $\delta_d^*$, $\epsilon^0=\tau/\max \{\rho_1^0, \rho_2^0\}$
\STATE \textbf{initialize:} $\bw^0$ 
\FOR {$k = 0,1,2,...$}


\STATE $j=0$\\

 \WHILE{$\Delta_d \leq \delta_d^*$ 
 or $r^j_r \leq \epsilon^k$ or $r^j_n \leq \epsilon^k$}
\STATE Calculate $\nabla_{\bw}\mathcal{L}_{\rho^k}({\bw}^j)$ using method in Appendix \ref{subsec:ParameterGradOpt} 
\STATE Initialize $\alpha_{j}$ with Barzilai-Borwein scheme
\STATE $\bw^{j+1} \;=\; {\cal P}_{[0,1]}\left(\bw^j \,-\, \alpha_{k}\nabla_{\bw}\mathcal{L}_{\rho^k}(\bw^j)\right)$ 
\WHILE{ $\mathcal{L}_{\rho^k}(\bw^{j+1})$ fails sufficient descent}
\STATE $\alpha_{k} = \alpha_{k}/2$
\STATE $\bw^{j+1} \;=\; {\cal P}_{[0,1]}\left(\bw^j \,-\, \alpha_{k}\nabla_{\bw}\mathcal{L}_{\rho^k}(\bw^j)\right)$
\ENDWHILE
\STATE compute $r_r^{j+1}$ and $r_n^{j+1}$, $j=j+1$
\ENDWHILE
\IF{$\Delta_p \leq \mathbf{\eta_{k}}$}
    \IF {$\Delta_p\leq \delta_p^*$ and $\Delta_d \leq \delta_d^*$}
        \STATE \textbf{stop} with approximate solution $\bw^{k+1}$
    \ENDIF \\
    \STATE $\mathbf{\lambda}^{k+1}_i = \lambda^k_i \,+\, \rho^k_i\, c_i^k $, $\rho_i^{k+1}=\rho_i^{k}$, \\
    \STATE $\eta_i^{k+1}=\eta^k/(\rho_i^k)^{0.9}$, $\rho_T=\max \{\rho_1,\rho_2 \}$ 

   \STATE $\epsilon^{k+1} = \epsilon^k/\rho_T$
\ELSE
\IF{$\Delta_p \leq \eta_{k,j}$}
   \STATE $\lambda_i^{k+1}=\lambda_i^{k}$, $\rho_i^{k+1}=2\rho_i^k$
   \STATE $\eta^{k+1}=1/(\rho_j^k)^{0.1}$, $\rho_T=\max \{\rho_1,\rho_2 \}$
   \STATE $\epsilon^{k+1} = \min\{1/\rho_T,\epsilon^k\}$
\ENDIF
\ENDIF

\ENDFOR    
\end{algorithmic}
\end{algorithm}

\section{Results}
\label{sec.Results}
\begin{table*}[t]
\begin{center}
\caption{$l_g^+$ and $\alpha/h$ for each library member}
\label{lgplus_comparison}
\begin{tabular}{|c|c|c|c|c|c|c|c|c|c|}
\hline
 Riblet Type & 1 & 2 & 3 & 4 & 5 & 6 & 7 & 8 & 9 \\
\hline
 $l_g^+$ & $10.8$ & $9.3$ & $11.1$ & $10.7$ & $11.2$ & $10.8$ & $9.8$ & $9.7$ & $10.2$\\
 \hline
 $\alpha/h$ & $0.057$ & $0.056$ & $0.051$ & $0.054$ & $0.055$ & $0.057$ & $0.056$ & $0.051$ & $0.053$\\
\hline
\end{tabular}
\end{center}
\end{table*}
In this section, we apply Algorithm~\ref{alg:cap} to determine the optimal shape of riblets with spanwise frequency $\omega_z=60$ on the lower wall of a turbulent channel flow with $Re_\tau=186$. The optimization algorithm is initialized from a selection of weights that satisfy $\mathds{1}^T\bw_r^0-1=0$ and ensure stability of the augmented dynamics ($\Gamma_{a_\mathrm{m}}\in\, {\cal C}$).  We exploit the spatial parameterization of our model by first conducting optimization based on the linearized dynamics of individual turbulent length scales, i.e., a single wavenumber pair $(k_x,\theta_z)$. We then compare results to those obtained based on multi-scale dynamics. Finally, given the nonconvexity of problem~\eqref{eq.problem-w}, we conduct a random seed experiment to analyze the dependence of solutions on initial conditions. The optimization algorithm was developed in MATLAB and was executed on a $4.50$ GHz AMD Ryzen 7 7700X 8-Core Processor with $32$ GB of RAM.

\begin{figure}
\hspace{0.3cm}

\begin{tabular}{cc}
	\hspace{-.6cm}
    \vspace{-0.3cm}
    \subfigure[]{\label{fig.EE_1_17_11_7}}\\
	\begin{tabular}{c}
	\vspace{.4cm}
	\rotatebox{90}{\footnotesize \text{rel. err. (\%)}}
	\end{tabular}
&
	\hspace{-.95cm}
	\begin{tabular}{c}
            \includegraphics[width=.43\textwidth]{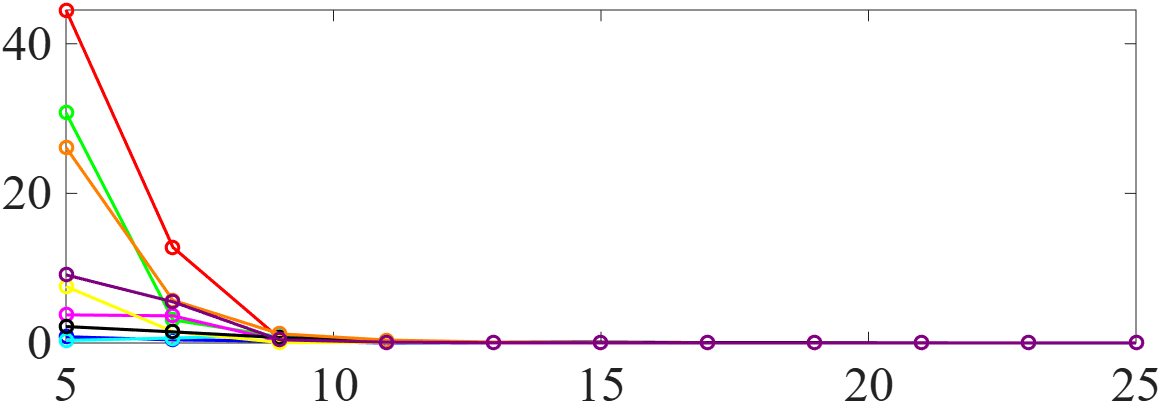}
		\\[-.1cm]
        
	\end{tabular}    
\\
	\hspace{-.6cm}
    \vspace{-0.3cm}
    \subfigure[]{\label{fig.EE_0_77_5_11}}\\
	\begin{tabular}{c}
	\vspace{.4cm}
	\rotatebox{90}{\footnotesize \text{rel. err. (\%)}}
	\end{tabular}
&
	\hspace{-.95cm}
	\begin{tabular}{c}
            \includegraphics[width=.43\textwidth]{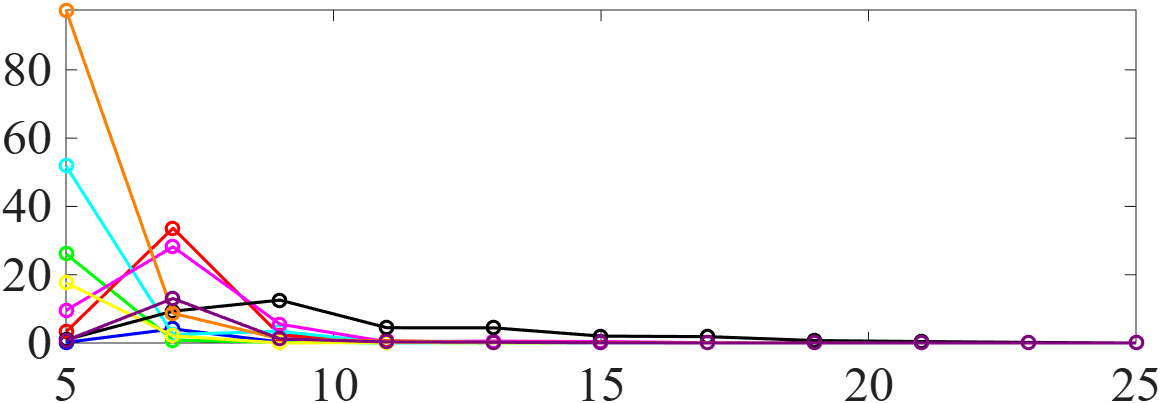}   
	\end{tabular}
\\
	\hspace{-.6cm}
    \vspace{-0.3cm}
    \subfigure[]{\label{fig.EE_8_72_1_28}}\\
	\begin{tabular}{c}
	\rotatebox{90}{\footnotesize \text{rel. err. (\%)}}
	\end{tabular}
&
	\hspace{-.95cm}
	\begin{tabular}{c}
            \includegraphics[width=.43\textwidth]{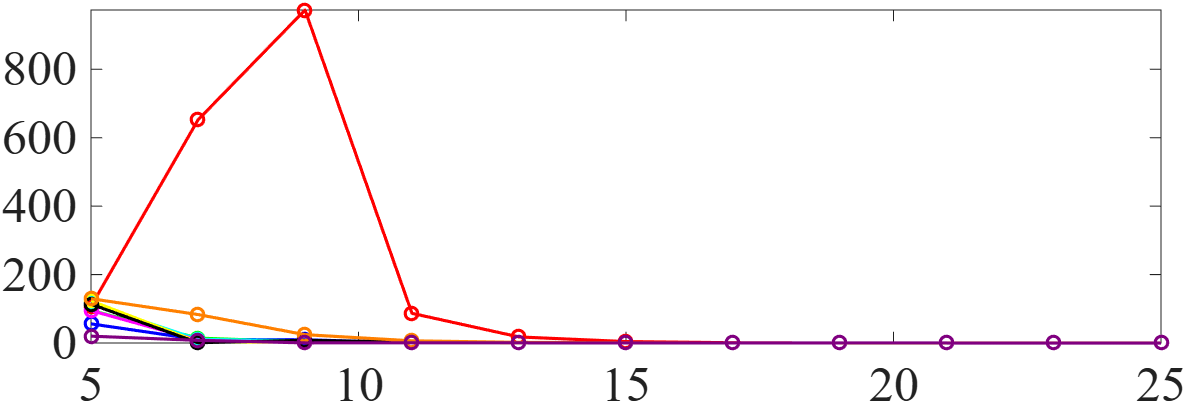}
		\\[-.1cm]
		{\hspace{0.2cm} \small $N_h$}
        \vspace{-0.3cm}
	\end{tabular}    
\end{tabular}
\vspace{0.1cm}
\caption{Convergence of energy by number of harmonics for riblets $(\textcolor{blue}{-})$ $1$, $(\textcolor{green}{-})$ $2$, $(\textcolor{red}{-})$ $3$, $(\textcolor{cyan}{-})$ $4$, $(\textcolor{magenta}{-})$ $5$, $(\textcolor{yellow}{-})$ $6$, $(-)$ $7$, $(\textcolor{orange}{-})$ $8$, and $(\textcolor{purple}{-})$ $9$ as identified in Fig. \ref{parametricCurves} for (a) $\bk_1=(1.17,11.7)$, (b) $\bk_2=(0.77,5.11)$ (c) $\bk_3=(8.28,1.28)$. rel. err. (\%) $\DefinedAs 100\times |1-E_{N_h} /E^*|$ where $E_{N_h}$ is the energy at any $N_h$ number of harmonics and $E^*$ corresponds to the energy with $25$ harmonics.}
\label{fig.ConvEnergy}
\end{figure}

\subsection{Riblet library synthesis}

Our shape-optimization framework relies on constructing a library of comparable riblet profiles that satisfy widely accepted geometric criteria for optimal drag reduction while maintaining a constant bulk flux. Specifically, we constrain the viscous-scaled groove size and the ratio of riblet height to channel half-height according to
\begin{equation}
9 \leq l_g^+ \leq 12,
\qquad
\frac{\alpha}{h} \leq 0.1,
\end{equation}
where $l_g^+ \DefinedAs \sqrt{A^+}$ and $A^+$ denotes the viscous-scaled groove area. The prescribed range of $l_g^+$ encompasses optimally drag-reducing riblets of various shapes~\cite{ranzarjovJFM21,garjim11b}. The height constraint limits interference with the buffer-layer viscous cycle and mitigates blockage effects~\cite{jim04,cas07}. Finally, to reduce the influence of secondary motions and spanwise dispersion, and thereby support the validity of linearization about $\bar{\bu} = [\,U(y,z)\,~0\,~0\,]^T$, we restrict attention to riblets with mid-height spacing ${\cal W}^+ \lesssim 20$, as suggested by~\cite{modendhutchu21}. Figure~\ref{parametricCurves} presents the library of riblet profiles with spacing $s^+=19.5$ considered in the optimization, with the corresponding geometric specifications summarized in Table~\ref{lgplus_comparison}.

\subsection{Optimization setup and performance}
\label{sec.singlescale}
Following the approach described in Section~\ref{sec.prob-formulation}, we partition the wall-normal domain into turbulent and laminar subregions located above and below $y=-1$, respectively. We discretize these subregions using $N_u=101$ collocation points above $y=-1$ and $N_b=20$ collocation points below it. Consistent with the methodology of~\cite{ranzarjovJFM21} for an optimal triangular riblet with a $75^\circ$ tip angle, we set the height of the laminar subregion ($r_p \alpha$) to $0.056\,\alpha$ and the magnitude of the resistance field ($R_0$ in Eq.~\eqref{Consistent_Pen}) to $3.9\times10^5$.

We retain the minimum number of harmonics, $N_h$, in the Fourier expansion of $K^{-1}(y,z)$ in Eq.~\eqref{eq.invK} required to accurately capture the energy of flow fluctuations over the riblets. Figure~\ref{fig.ConvEnergy} shows the relative error introduced by harmonic truncation across several turbulent length scales. Based on these results, we use $N_h=15$ for the riblet-optimization computations presented in the remainder of this section and employ a higher-fidelity model with $N_h=25$ to evaluate the flow physics associated with the optimized riblets. 

\begin{figure}
\vspace{1.4cm}
\begin{tabular}{rcc}
	\hspace{-.6cm}
	\begin{tabular}{c}
	\vspace{.4cm}
	\rotatebox{90}{\footnotesize $\norm{\bw^k-\bw^\star}_2/\norm{\bw^\star}_2$}
	\end{tabular}
&
	\hspace{-.95cm}
	\begin{tabular}{c}
            \includegraphics[width=.225\textwidth]{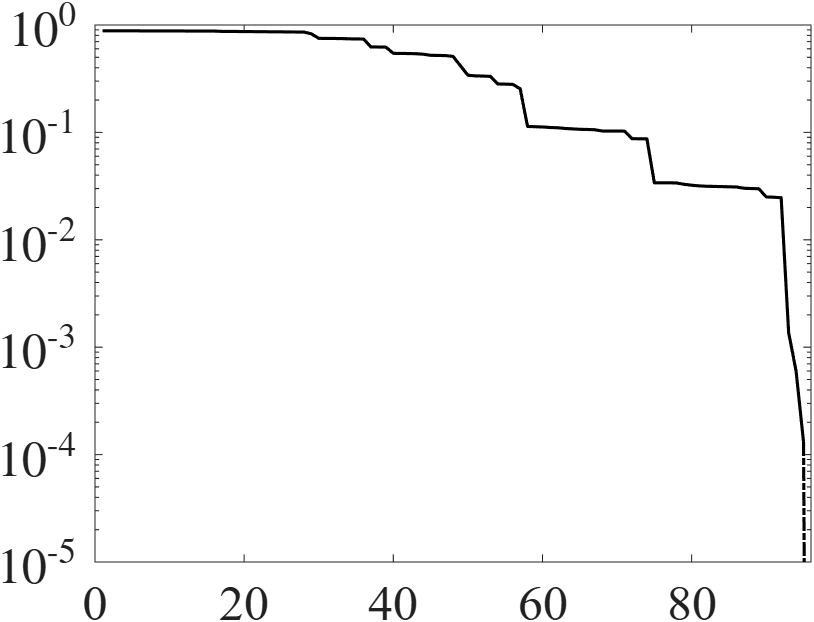}
		\\[-.1cm]
		{\small iteration}
	\end{tabular}
&
	\hspace{-0.7cm}
	\begin{tabular}{c}
         	\includegraphics[width=.225\textwidth]{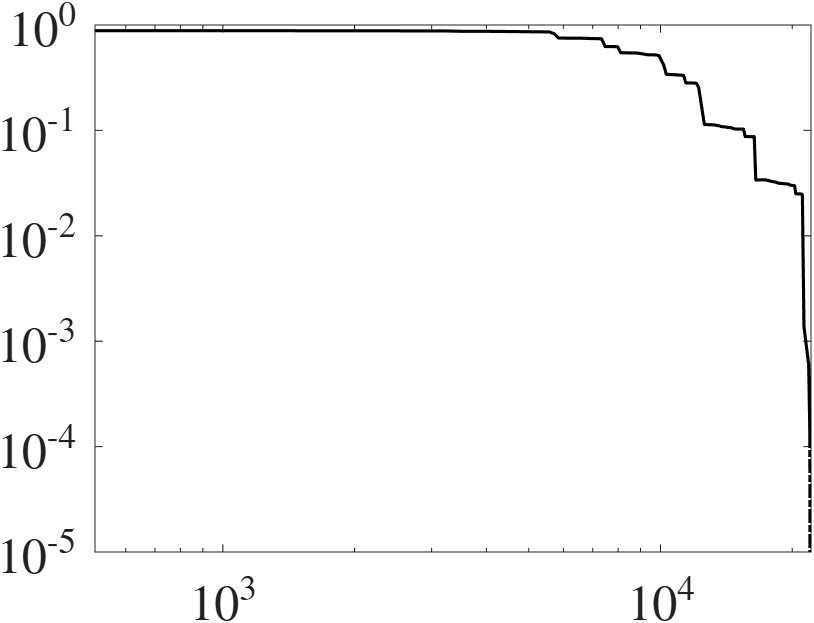}
		\\[-.1cm]
		{\small solve time (sec)}
	\end{tabular}
	\\[-.2cm]
&
	\hspace{-1cm}
	\subfigure[]{\label{fig.iter-MM}}
&
	\hspace{-1cm}
	\subfigure[]{\label{fig.time-MM}}
\end{tabular}
\vspace{-.4cm}
\caption{ Convergence curves of Algorithm~\ref{alg:cap} ($-$) versus (a) iteration number; and (b) cumulative solve time for a single length scale optimization with $(k_x,\theta) = (1.17,11.7)$. Here, $\bw^\star$ denotes the final iterate.}
\label{fig.convergence-MM}
\end{figure}

Figure~\ref{fig.convergence-MM} shows the convergence histories of the customized MM algorithm introduced in Sec.~\ref{sec.alg} when applied to problem~\eqref{eq.problem-w} at $(k_x,\theta_z)=(1.17,11.7)$. This wavenumber pair corresponds to the characteristic length scales of streamwise-elongated streaks that dominate the near-wall cycle. The MM algorithm was initialized from a riblet profile resulting from an equal distribution of weights among library members (Fig.~\ref{fig.eqWeightRiblet1}). The primal and dual stopping tolerances are set to $\delta^*_p=[10^{-4},4.36\times10^{-4}]$ and $\delta^*_d=10^{-4}$, respectively. To assess the asymptotic convergence behavior, we estimate the rate $q$ from
\begin{align*}
    ||\bw^{k+1}-\bw^*||^2_2 \leq q ||\bw^{k}-\bw^*||^2_2
\end{align*}
for sufficiently large $k$. Here $\bw^*$ is the value obtained at final iteration. The resulting values do not satisfy $q\in(0,1)$ for sufficiently large $k$, indicating sub-linear convergence.

\subsection{Multi-scale optimization}
\label{sec.multiscale}

\begin{figure}[h]
\hspace{0.3cm}
\begin{tabular}{rcc}
	\hspace{-.8cm}
	\begin{tabular}{c}
	\vspace{.4cm}
	\rotatebox{90}{$w_i$}
    \hspace{.1cm}
	\end{tabular}
&
	\hspace{-.95cm}
	\begin{tabular}{c}
            \includegraphics[width=.43\textwidth]{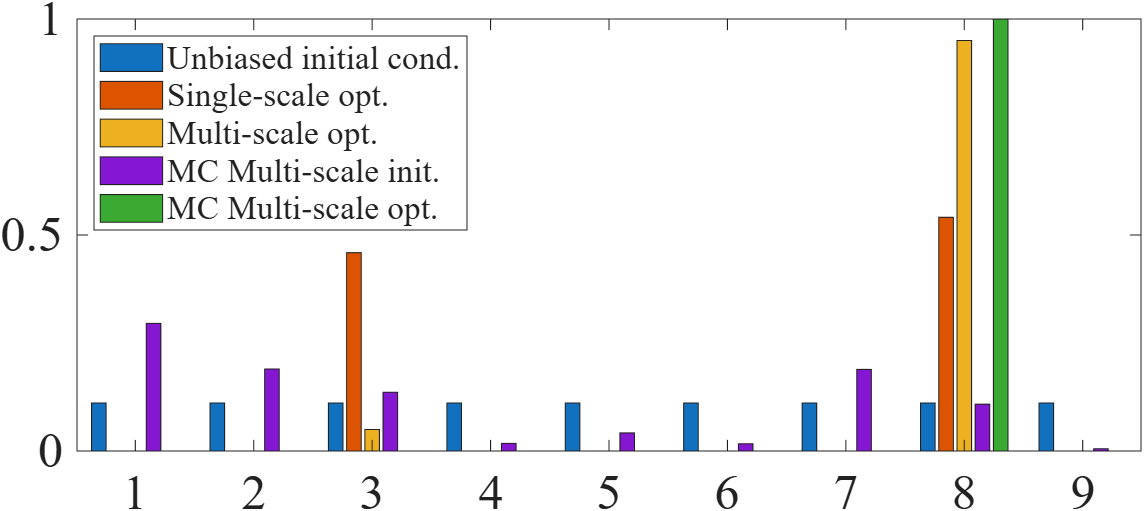}
		\\
		{riblet index}
	\end{tabular}
\end{tabular}
\caption{Optimal weight magnitudes resulting from single-scale and multi-scale optimization, in addition to the best result of the random seed optimization. The initial weights for each optimization has also been included.}
\label{fig.singleScaleFailing}
\end{figure}

\begin{figure}[t]
\vspace{.2cm}
\begin{tabular}{ccrc}
\subfigure[]{\label{fig.eqWeightRiblet1}}
\hspace{0.2cm}
\vspace{-0.2cm}
&
\hspace{3cm}
\subfigure[]{\label{fig.optRiblet1}}
\\
	\hspace{0.5cm}
	\begin{tabular}{c}
	\vspace{.4cm}
    \hspace{-1.7cm}
	\rotatebox{90}{\footnotesize $y$}
	\end{tabular}
&
	\hspace{-1.35cm}
	\begin{tabular}{c}
            \includegraphics[width=.23\textwidth]{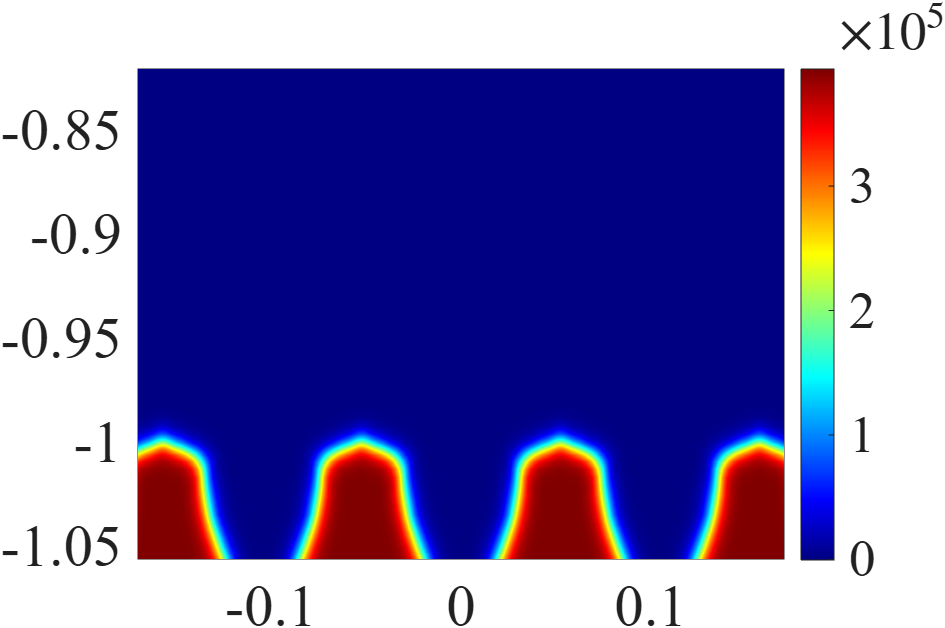}
		\\[-.1cm]
		{\small $z$}
	\end{tabular}
&
	\hspace{-0.9cm}
	\begin{tabular}{c}
            \includegraphics[width=.23\textwidth]{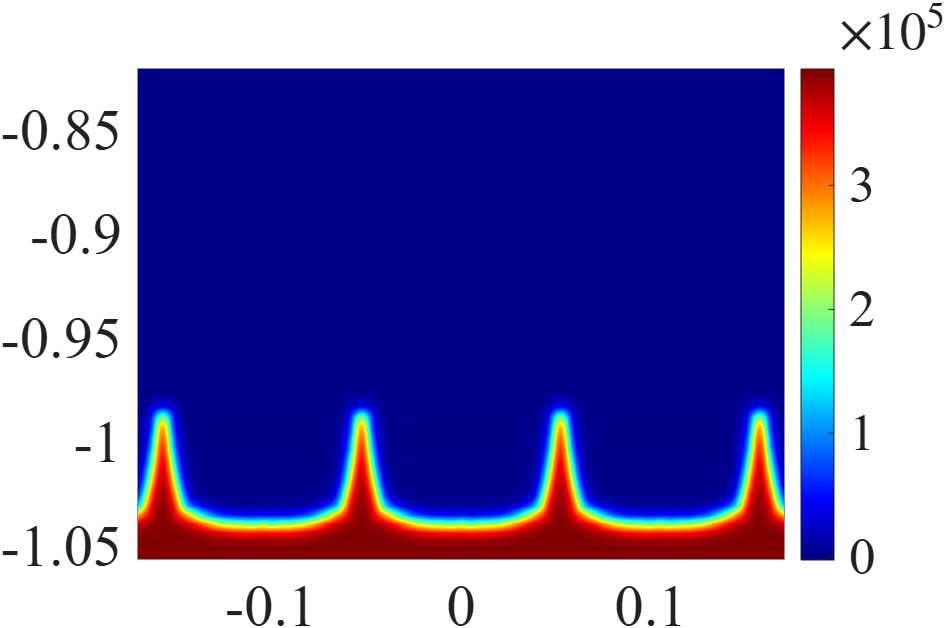}
		\\[-.1cm]
		{\small $z$}
	\end{tabular}
	\\[-.1cm]
\end{tabular}
\\
\begin{tabular}{rcrc}
\subfigure[]{\label{fig.eqWeightRiblet2}}
\hspace{0.2cm}
\vspace{-0.2cm}
&
\hspace{3cm}
\subfigure[]{\label{fig.optRiblet2}}
\\
	\hspace{0.5cm}
	\begin{tabular}{c}
	\vspace{.4cm}
    \hspace{-1.7cm}
	\rotatebox{90}{\footnotesize $y$}
	\end{tabular}
&
	\hspace{-1.35cm} 
	\begin{tabular}{c}
            \includegraphics[width=.23\textwidth]{initialRibletFinal.png}
		\\[-.1cm]
		{\small $z$}
	\end{tabular}
&
	\hspace{-0.9cm}
	\begin{tabular}{c}
            \includegraphics[width=.23\textwidth]{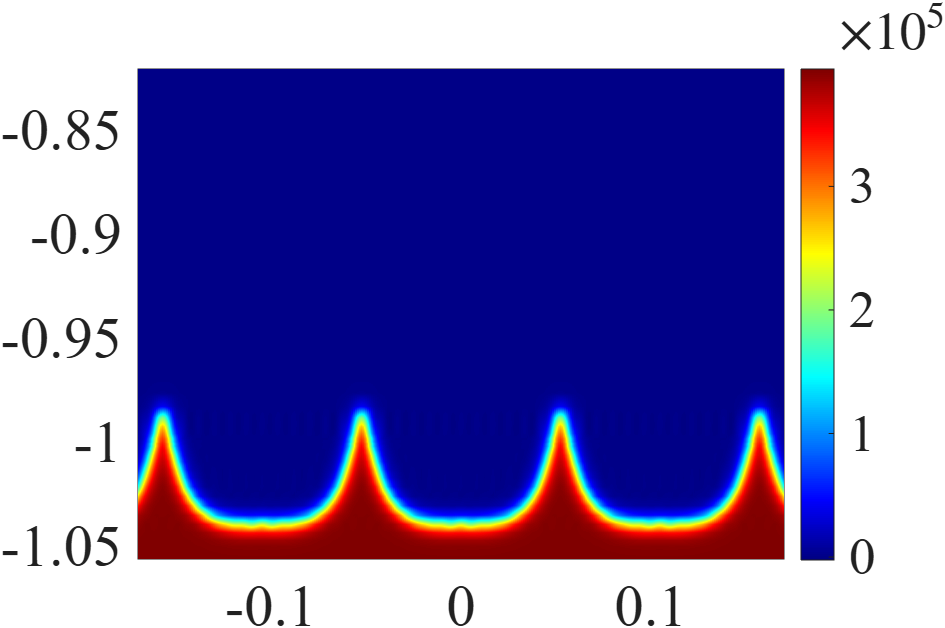}
		\\[-.1cm]
		{\small $z$}
	\end{tabular}
	\\[-.1cm]
\end{tabular}
\\

\begin{tabular}{rcrc}
\subfigure[]{\label{fig.eqWeightRiblet3}}
\hspace{0.2cm}
\vspace{-0.2cm}
&
\hspace{3cm}
\subfigure[]{\label{fig.optRiblet3}}
\\
	\hspace{0.5cm}
	\begin{tabular}{c}
	\vspace{.4cm}
    \hspace{-1.7cm}
	\rotatebox{90}{\footnotesize $y$}
	\end{tabular}
&
	\hspace{-1.35cm}
	\begin{tabular}{c}
            \includegraphics[width=.235\textwidth]{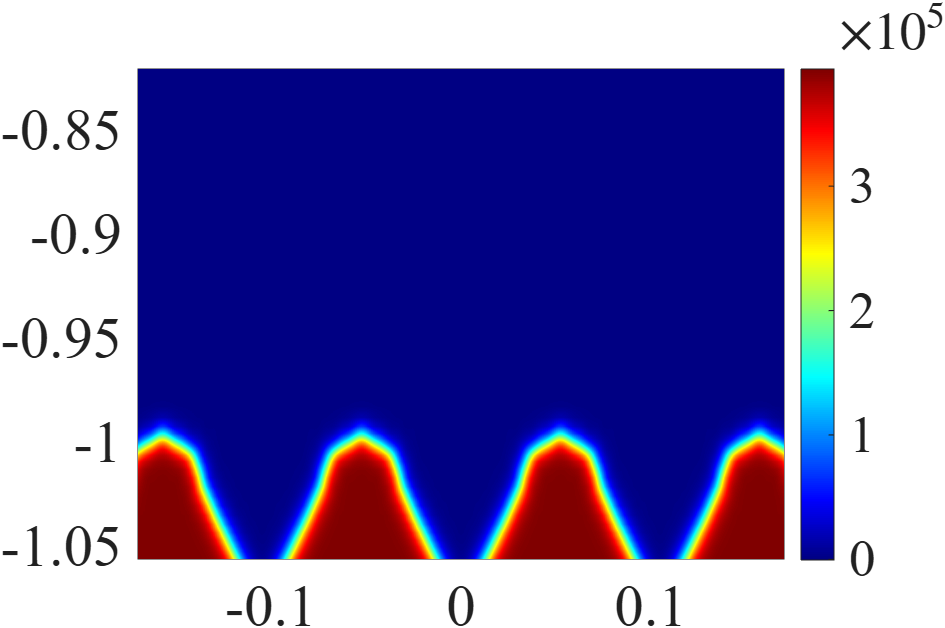}
		\\[-.1cm]            
		{\small $z$}
	\end{tabular}
&
	\hspace{-0.9cm}
	\begin{tabular}{c}
            \includegraphics[width=.235\textwidth]{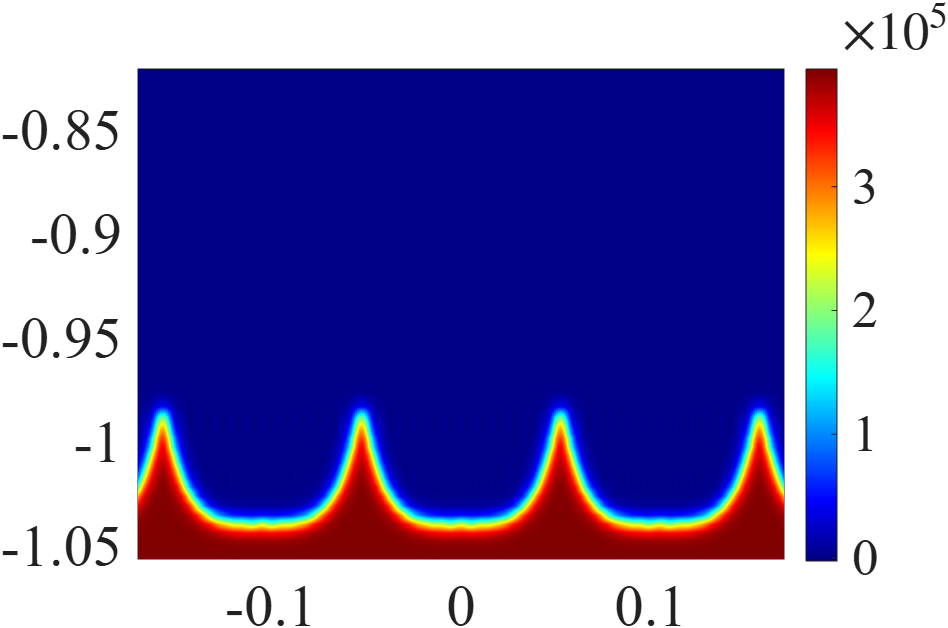}
		\\[-.1cm]            
		{\small $z$}
	\end{tabular}
\end{tabular}
\vspace{-.35cm}
\caption{Resistance function $K^{-1}(y,z)$ for  $\omega_z=60$ corresponding to the initial weight distribution (left column) and final solution of problem~\eqref{eq.problem-w} (right column). (a) single-scale initial condition; (b) single-scale solution; (c) multi-scale initial condition; (d) multi-scale solution; (e) the initial condition for the best result of the random seed optimization; (f) the best result of the random seed optimization. Solutions were obtained with the optimal tolerances $\delta_p^*$ and $\delta_d^*$ described in subsection~\ref{sec.multiscale}.}
\label{fig.optimalRibletShapes}
\end{figure}

Figure~\ref{fig.singleScaleFailing} compares the weight distributions obtained from shape optimization using single-scale and multi-scale cost functions. As evident from this figure and Fig.~\ref{fig.optimalRibletShapes}, although the single-scale optimization targets the dynamically dominant structures associated with the near-wall cycle, accounting for the energy costs of two additional dynamically significant modes leads to a different distribution of weights and, consequently, to different optimal riblet profiles. Specifically, the multi-scale formulation includes the energy cost associated with an energetically amplified mode, $(k_x,\theta_z)=(8.7182,1.28)$, and the most strongly suppressed mode, $(k_x,\theta_z)=(0.77,5.11)$, resulting from the introduction of triangular riblets with a $75^\circ$ tip angle. Although the resulting geometric differences are relatively small, they produce noticeable changes in the achievable energy suppression and drag reduction; see Table~\ref{ResultsSummary}. The skin-friction drag associated with the ribleted surfaces is computed using the turbulence-modeling framework of Ran et al.~\cite{ranzarjovJFM21}, in which riblet-induced modifications to the mean flow are obtained from the corrections to the turbulent eddy viscosity in Eqs.~\eqref{eq.meanflow}.

The energy suppression and drag reduction values reported in Table~\ref{ResultsSummary} are computed using a higher-fidelity model of the flow over the optimized riblets, with $N_h=25$ and spectral ranges $k_x\in[-1.5,1.6]$ and $\theta_z\in[0,30]$. These results demonstrate the importance of accounting for a broader range of dynamically relevant scales. In particular, optimizing the riblet shape for a single dominant scale does not necessarily yield improved overall performance. Finally, while the resulting performance is comparable to that reported for high-performance riblets in the existing literature, the riblet profiles identified by our optimization framework provide a substantial improvement.

\subsection{Random-seed experiments}
\label{sec.randomseed}

\begin{figure}[t]
\hspace{-.2cm}
\begin{tabular}{cc}
    \subfigure[]{\label{fig.MC_Weights}}
    &
    \\[-.15cm]
	\begin{tabular}{c}
	\vspace{.4cm}
	\rotatebox{90}{seed number}
	\end{tabular}
    &
    \hspace{-0.6cm}
	\begin{tabular}{c}
            \includegraphics[width=0.38\textwidth]{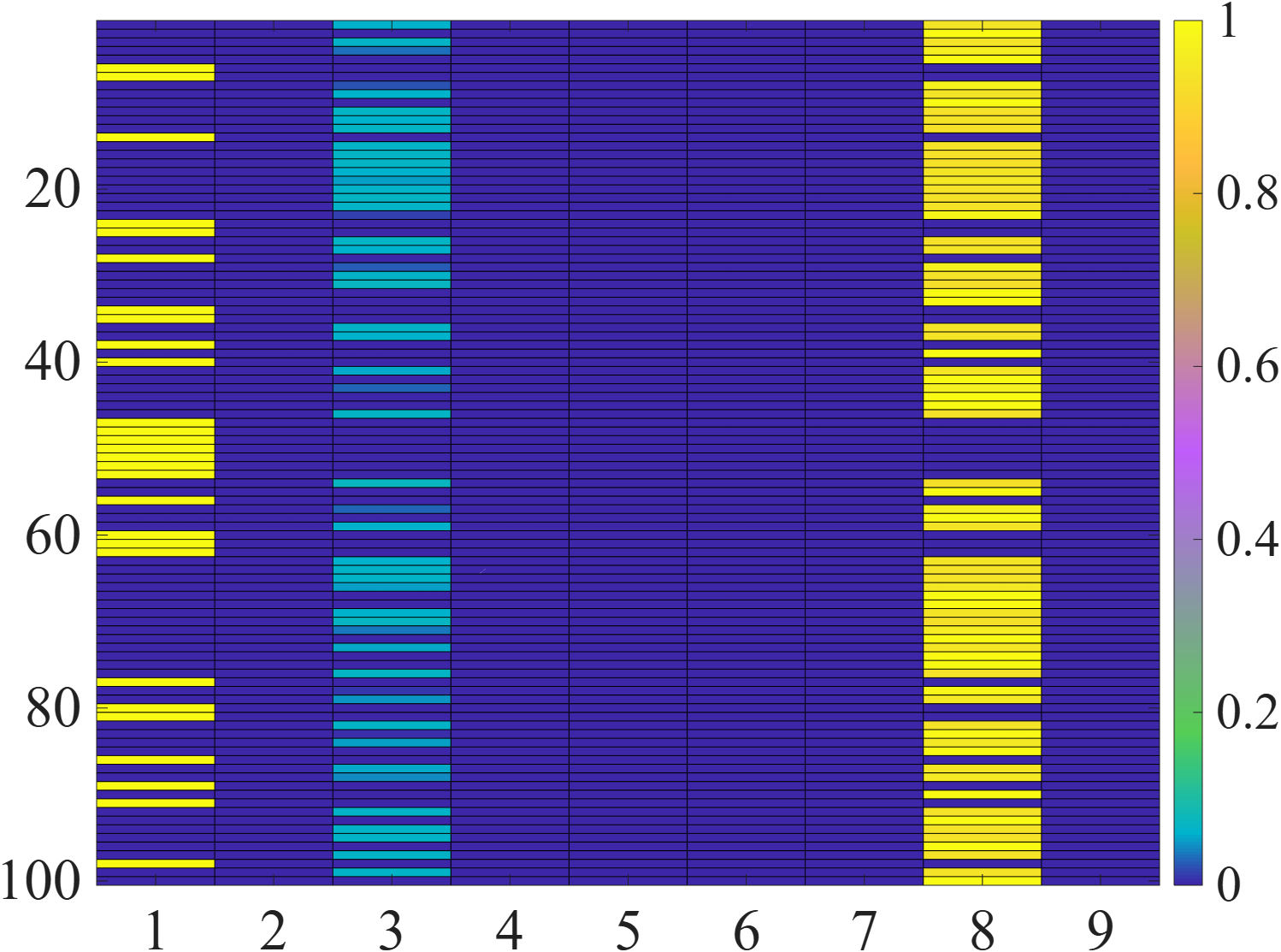}
		\\[-.05cm]
		{riblet index}
	\end{tabular}
    \\
    \subfigure[]{\label{fig.MC_Cost}}   
    &
    \\
    \begin{tabular}{c}
	\vspace{.4cm}
	\rotatebox{90}{$J(\bw)$}
	\end{tabular}
    &
    \hspace{-.6cm}
	\begin{tabular}{c}
            \includegraphics[width=.38\textwidth]{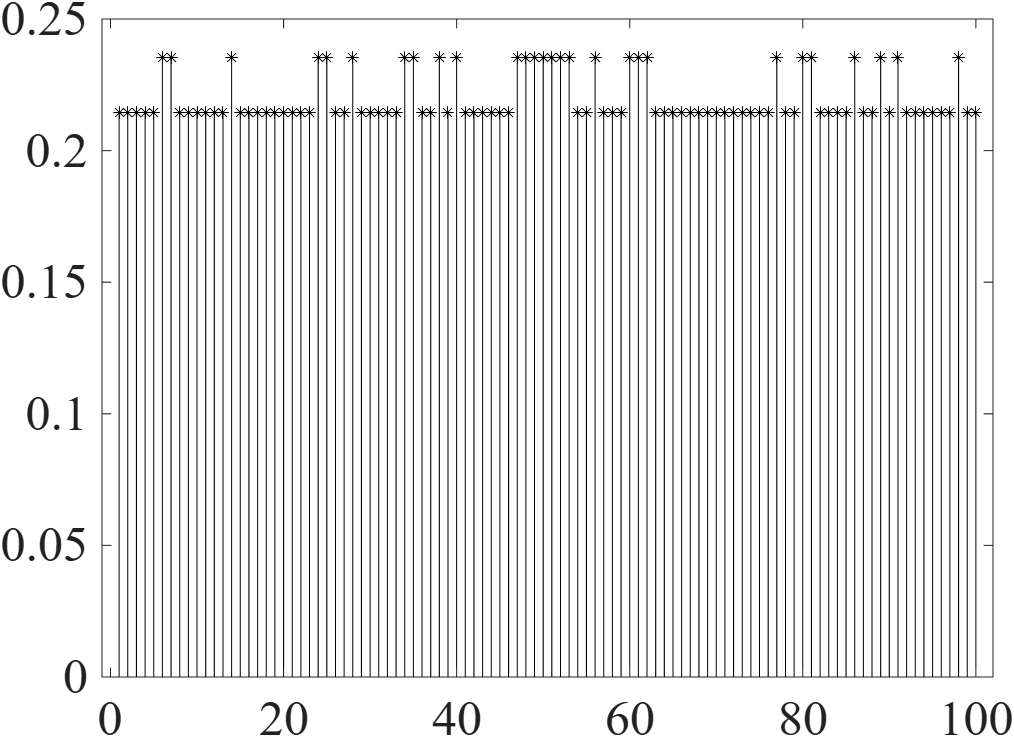}
		\\[-.05cm]
		{seed number}
	\end{tabular}
\end{tabular}
\caption{Randomized initialization results. (a) Distribution of optimal weights $\bw^\star$ vs seed number; and (b) the cost functional $J(\bw)$ by seed index number.}
\label{fig.MC-Visualization}
\end{figure}

Due to the nonconvexity of problem~\eqref{eq.problem-w}, solutions obtained using the MM algorithm can depend on the initial condition. To elucidate this dependence, we conduct a random-seed experiment in which multi-scale optimization is conducted from 100 different initial weight distributions. Figure~\ref{fig.MC-Visualization} shows the final distribution of weights among the 9-member library in addition to the final value of the objective function resulting from each initial seed. We observe a variety of local optima. However, we observe a clustering of results around indices 3 and 8, which correspond to blade- and scallop-shaped members of our library, respectively. 

The best result obtained from this experiment is compared against the single-scale and multi-scale computations presented in the previous subsection, which were initialized from an initial condition that results from an equal distribution of weights. As seen in Table~\ref{ResultsSummary}, the riblet shape resulting from the best solution of the random-seed experiment corresponds to a scallop-shaped riblet and outperforms both single-scale and multi-scale minimizers generated from deterministic initial conditions. Critically, the $l_g^+$ of all three shapes falls within the previously established optimal range of $l_g^+\in[9.7,11.7]$ ~\cite{ranzarjovJFM21,garjim11b}. Importantly, small variations in the energy of fluctuations results in meaningful improvements in drag reduction when post-processing calculations across the full spectrum of horizontal scales.

\begin{table}[H]
\begin{center}
\caption{Comparison of energy suppression and drag reduction resulting from the optimal riblet shapes obtained from single-scale, multi-scale, and random seed optimizations.}
\label{ResultsSummary}
\begin{tabular}{|c|c|c|c|c|}
\hline
 & $\Delta E_{total}$ (\%) & $\Delta D$ (\%) & $l_g^+$ & $\alpha/h$ \\
\hline
Triangular \cite{ranzarjovJFM21} & $-5.44$ & $-8.42$ & $11.7$ & $0.066$\\
\hline
Single-scale & $-5.61$ & $-8.59$ & $10.26$ & $0.051$\\
\hline
Multi-scale & $-5.87$ & $-9.09$ & $9.77$ & $0.051$\\
\hline
Randomized  & $-5.89$ & $-9.12$ & $9.72$ & $0.051$\\
\hline
\end{tabular}
\end{center}
\end{table}

\section{Concluding Remarks}
\label{sec.conclusions}

In this study, we developed a model-based framework for optimizing riblet geometries by examining their influence on the energy of flow fluctuations. The framework formulates riblet design as a constrained $\mathcal{H}_2$ optimal control problem, enabled by the evolution form of the linearized Navier–Stokes equations and the use of volume penalization to represent the effects of riblets on the flow dynamics. Signed-distance functions were employed to parameterize the riblet geometries, allowing the original feedback-gain optimization problem to be reformulated as a weight-selection problem. We then developed a customized algorithm to solve the resulting optimization problem. Although the random-seed optimization results suggest that scallop-shaped riblets may be optimal within the considered design space, we also showed that appropriately weighted combinations of blade- and scallop-shaped riblets constitute competitive alternatives.

The present approach uses the energy of flow fluctuations as a surrogate measure for drag reduction. Further analysis is therefore needed to determine whether the identified riblet geometries are also optimal with respect to skin-friction drag. To reduce the computational cost, the optimization was restricted to a prescribed riblet spacing, and the geometry library was limited to nine commonly studied riblet shapes from the literature. Future work should expand this library to include broader variations in riblet height, width, and spacing, particularly geometries that lie outside the convex hull defined by the current parameterization.

The proposed low-complexity modeling framework relies on linearization about a base-flow profile containing only a streamwise velocity component. More sophisticated models that account for dispersive flow motions could provide more accurate predictions of riblet performance and enable the consideration of additional geometric families that are currently excluded by the modeling assumptions. Ultimately, a comprehensive evaluation of the optimal riblet geometries will require verification of the model-based predictions through high-fidelity numerical simulations.

\appendix

\subsection{$\Gamma_{a_\mathrm{m}}$ in the decomposition of the state matrix $A(\bk)$}
\label{subsec:strcgamfeedmat}

The structured bi-infinite matrix $\Gamma_{a_\mathrm{m}}$,
\begin{align*}
{\small
	\Gamma_{a_\mathrm{m}}
	~\DefinedAs~
	\left[
	\ba{ccccc}
         \ddots & \vdots & \vdots & \vdots & \rotatebox{90}{$\ddots$}
         \\
         \cdots & \Gamma_{\mrn-1,0} & \Gamma_{\mrn-1,+1} & \Gamma_{\mrn-1,+2} & \cdots
         \\[0.1cm]
         \cdots & \Gamma_{\mrn,-1} & \Gamma_{\mrn,0} & \Gamma_{\mrn,+1} &  \cdots
         \\[0.1cm]
         \cdots &  \Gamma_{\mrn+1,-2} & \Gamma_{\mrn+1,-1} & \Gamma_{\mrn+1,0} & \cdots
         \\
         \rotatebox{90}{$\ddots$} & \vdots & \vdots & \vdots & \ddots
    	\ea
	\right]
}
\end{align*}
represents riblet-induced changes to the dynamics of velocity fluctuations with operator $\Gamma_{\mrn,\mrm}$ representing the influence of the ($n+m$)th harmonic of the state $\hat{\bpsi}_{n+m}$ on the dynamics of the $n$th harmonic $\hat{\bpsi}_{n}$. These operators themselves have the block structure
\begin{align}
    \Gamma_{\mrn,\mrm} \;=\; \tbt{\Gamma_{11}}{\Gamma_{12}}{\Gamma_{21}}{\Gamma_{22}}
\end{align}
where 
\begin{align*}
{\small
\hspace{-.5cm}
        \ba{rcl}
        \Gamma_{1,1}
        &\!\!\!\!\!\!\!\!=\!\!\!\!\!\!\!\!\!&
        \Delta_n^{-1} \!\! \left[2\, \mri m\, k_x\, \omega_z\dfrac{\theta_{n+m}}{k_{n+m}^2}\left(U'_{-m}\partial_y +\, U_{-m}\partial_{yy}\right)\right.
        \\[0.4cm]
        &&
        -\, a_{-m}\Delta_{n+m}
        +\,\mri k_x(m\, \omega_z)^2U_{-m} -\, a'_{-m}\partial_y
        \\[0.3cm]
        &&
        - 2\, m\, k_x \omega_z \theta_{n+m}U_{-m} +\, m\, \omega_z (m\omega_z - 2\, \theta_{n+m})a_{-m} 
        \\[0.3cm]
        &&
        + \left. m\, \omega_z\dfrac{\theta_{n+m}}{k_{n+m}^2}\left(a'_{-m}\partial_y\,+\,a_{-m}\partial_{yy}\right)\right]
\\[0.45cm]
        \Gamma_{1,2}
        &\!\!\!\!\!\!\!\!=\!\!\!\!\!\!\!\!\!&
        \Delta_n^{-1}\left[2\,\dfrac{\mri\, m\, k_x^2\,\omega_z}{k_{n+m}^2}\left(U'_{-m}\,+\,U_{-m}\partial_y\right)\right.
        \\[0.35cm]
        && +\,\left.\dfrac{m\, k_x\,\omega_z}{k_{n+m}^2}\left(a'_{-m} +\, a_{-m}\partial_y\right)\right]
\\[0.45cm]
        \Gamma_{2,1}
        &\!\!\!\!\!\!\!\!=\!\!\!\!\!\!\!\!\!&
        \mri\, m\, \omega_z \left(U'_{-m} \,-\, U_{-m}\partial_y\right) 
        \\[0.25cm]
        && + \left[\mri(m\,\omega_z)^2\dfrac{\theta_{n+m}}{k_{n+m}^2}\,U'_{-m} - \dfrac{m\, k_x\, \omega_z}{k_{n+m}^2}\,a_{-m} \right]\partial_y
        \\[0.45cm]
        \Gamma_{2,2}
        &\!\!\!\!\!\!\!\!=\!\!\!\!\!\!\!\!\!&
        \Big[\big(\dfrac{m\,\omega_z}{k_{n+m}}\big)^2 - 1\Big] \mri k_x U_{-m}
        + 
        \Big[m \omega_z \dfrac{\theta_{n+m}}{k_{n+m}^2} -1 \Big] a_{-m}
        \ea
}
\end{align*}
Here, $\theta_{n+m} = (n\,+\,m)\omega_z \,+\, \theta$, $k_{n+m}^2 = k_x^2 \,+\, \theta_{n+m}^2$,  $\Delta_{n+m}=\partial_{yy} \,-\, k_{n+m}^2$, and $U_{-m}=\diag\left(\hat{U}_{-m}\right)$.

\subsection{Computation of the gradient $\nabla J(\bw)$}
\label{subsec:ParameterGradOpt}

As shown in Sec.~\ref{sec.prob-formulation}, the objective function contains an integration over all scales $\bk$. This can be discretized such that the objective function of problem~\eqref{eq.problem-w} can be expressed as
\begin{align}
\label{eq.newcost}
    J(\Gamma_{a_\mathrm{m}})
    \;=\; 
    \ds{\sum_{j=1}^{N_\bk}}\trace{\left(X_j\, P_{M,j}\right)}
\end{align}
where $N_\bk$ is the number of scales considered in approximating the cost function and the covariance corresponding to the $j$th scale, $X_j$, solves the Lyapunov equation
\begin{align}
    \label{eq.lyap1}
    \ba{l}
    \left(A_{s,j}+\Gamma_{a_\mathrm{m},j}\right)^* X_j \,+\, X_j\left(A_{s,j}+\Gamma_{a_\mathrm{m},j}\right) \,+\,
    Q_j \,+\, \; 
    \\[.15cm]
    \hspace{4.5cm}
    \Gamma_{a_\mathrm{m},j}^*\, R_j\,\Gamma_{a_\mathrm{m},j} \;=\; 0
    \ea
\end{align}
where $P_{M,j} \DefinedAs BMB^*$, and $\Gamma_{a_\mathrm{m},j}$ are defined over the set of stabilizing augmentation matrices for $A_s$, i.e., $\Gamma_{a_\mathrm{m},j}\in {\cal C}$. Because the objective gradients associated with individual scales can be computed independently, the total gradient can be evaluated in parallel across the $N_{\bk}$ scales. For notational simplicity, throughout the remainder of this appendix we omit the subscript $j$, which identifies the particular $\bk$-pair associated with each dynamical system.

As shown in~\cite[Lemma 3.1]{rau97}, the objective function $J$ is indeed differentiable over the set of stabilizing gain matrices.
Building on ~\cite[Theorem 3.2]{rau97}, the following theorem provides an expression for the derivative of the objective function $J$ with respect to weighting parameters $w_i$.

 \vsp

\begin{theorem}
    \label{thm.gradient}
    The derivative of the objective function $J(\bw)$ with respect to the $i$th entry of $\bw$ is given by $\trace \left(2L (\tfrac{\partial\Gamma_{a_\mathrm{m}}}{\partial w_i}) ^* (X+R\Gamma_{a_\mathrm{m}})\right)$, where $L \in \bbC^{N_\bw\times N_\bw}$ solves the Lyapunov equation
    \begin{align*}
 L\left(A_s \,+\,\Gamma_{a_\mathrm{m}}\right)^* \,+\, \left(A_s\,+\,\Gamma_{a_\mathrm{m}}\right)L \;+\; P_M \;=\;0.
 \end{align*}
\end{theorem}
 
 \textit{Proof:} 
The derivative of the objective function $J(\bw)$ with respect to the weight $w_i$ is given by 

\begin{align}
 \label{eq.XDerivative}
 \dfrac{\partial J(\bw)}{\partial w_i}
 \,=\,
\trace{\left(\dfrac{\partial X}{\partial \Gamma_{a_\mathrm{m}}} \dfrac{\partial \Gamma_{a_\mathrm{m}}}{\partial w_i}\,P_M \right)}.
 \end{align}
Let
 \begin{align*}
 \Psi(X,\Gamma_{a_\mathrm{m}}) \;\DefinedAs\; (A_s+\Gamma_{a_\mathrm{m}})^* X \,+\, X(A_s+\Gamma_{a_\mathrm{m}}) \,+ 
 \\[.15cm]
Q\,+\,\Gamma_{a_\mathrm{m}}^* R\Gamma_{a_\mathrm{m}}.
 \end{align*}
The function $\Psi$ and its derivative are zero, i.e.,
\begin{align}
\label{eq.Psi-diff}
\Psi_X(X,\Gamma_{a_\mathrm{m}})\,\dfrac{\partial X}{\partial w_i} 
\,+\, 
\Psi_{\Gamma}(X,\Gamma_{a_\mathrm{m}})\,\dfrac{\partial \Gamma_{a_\mathrm{m}}}{\partial w_i}
\;=\;
0
\end{align}
which yields 
\begin{align}
\label{eq.diffDerive}
\dfrac{\partial X}{\partial \Gamma_{a_\mathrm{m}}} \dfrac{\partial \Gamma_{a_\mathrm{m}}}{\partial w_i} (A_s+\Gamma_{a_\mathrm{m}}) \,+\, (A_s + \Gamma_{a_\mathrm{m}})^*\dfrac{\partial X}{\partial \Gamma_{a_\mathrm{m}}} \dfrac{\partial \Gamma_{a_\mathrm{m}}}{\partial w_i} \;+ ~~~~
&
\non 
\\[.15cm] 
\left(\dfrac{\partial \Gamma_{a_\mathrm{m}}}{\partial w_i}\right)^*\!(X+R\Gamma_{a_\mathrm{m}}) \,+\,(X+R\Gamma_{a_\mathrm{m}})^*\left(\dfrac{\partial \Gamma_{a_\mathrm{m}}}{\partial w_i}\right)  &\,=\;  0.
\end{align}
This equation is obtained by substituting the partial derivatives of $\Psi$, which are denoted by the subscripts, i.e., 

\begin{align*}
    \Psi_{\Gamma}(X,\Gamma_{a_\mathrm{m}}) \dfrac{\partial\Gamma_{a_\mathrm{m}}}{\partial w_i} 
    &\,=\,
    X \dfrac{\partial\Gamma_{a_\mathrm{m}}}{\partial w_i} \, +\, \left(\dfrac{\partial\Gamma_{a_\mathrm{m}}}{\partial w_i} \right)^*\!X \;+  
    \\[.15cm]
    &
    \hspace{1.cm}\left(\dfrac{\partial\Gamma_{a_\mathrm{m}}}{\partial w_i} \right)^*\!R \Gamma_{a_\mathrm{m}} \,+\, \Gamma_{a_\mathrm{m}}^* R \dfrac{\partial\Gamma_{a_\mathrm{m}}}{\partial w_i} 
\end{align*}
and
\begin{align*}    
    \Psi_X(X,\Gamma_{a_\mathrm{m}}) \dfrac{\partial X}{\partial w_i} 
    &\,=\,
    (A_s+\Gamma_{a_\mathrm{m}})^* \dfrac{\partial X}{\partial \Gamma_{a_\mathrm{m}}}\dfrac{\partial \Gamma_{a_\mathrm{m}}}{\partial w_i} \;+
    \\[.15cm]
    &
    \hspace{2.5cm} \dfrac{\partial X}{\partial \Gamma_{a_\mathrm{m}}}\dfrac{\partial \Gamma_{a_\mathrm{m}}}{\partial w_i}(A_s\,+\,\Gamma_{a_\mathrm{m}})
\end{align*}
into Eq.~\eqref{eq.Psi-diff}.
Multiplying Eq.~\eqref{eq.diffDerive} by $L$ from the left and the dual of the Lyapunov equation~\eqref{eq.lyap1}, i.e., 
\begin{align}
\label{eq.ExtraLyap}
 L \left(A_s+\Gamma_{a_\mathrm{m}}\right)^* \,+\,\left(A_s+\Gamma_{a_\mathrm{m}}\right)L \,+\, P_M \;=\; 0,
\end{align}
by $\dfrac{\partial X}{\partial \Gamma_{a_\mathrm{m}}}\dfrac{\partial \Gamma_{a_\mathrm{m}}}{\partial w_i}$ from the left and taking the trace of both sides of each equation yields the identity
\begin{align*}
~\trace \left(2L \left(\dfrac{\partial \Gamma_{a_\mathrm{m}}}{\partial w_i}\right)^* \!\left(X + R \Gamma_{a_\mathrm{m}}\right)\right) = &
\\[.15cm]
&
\hspace{-.8cm}\trace\left(\dfrac{\partial X}{\partial \Gamma_{a_\mathrm{m}}}\dfrac{\partial \Gamma_{a_\mathrm{m}}}{\partial w_i} P_M \right)
\end{align*}
Comparing the right-hand-side of this identity with that of Eq.~\eqref{eq.XDerivative} yields 
\begin{align}
\dfrac{\partial J (\bw)}{\partial w_i} 
\,=\,
\trace{\left(2L \,\left(\dfrac{\partial\Gamma_{a_\mathrm{m}}}{\partial w_i}\right) ^* \left(X + R\Gamma_{a_\mathrm{m}}\right)\right)}
\end{align}
and completes the proof.

In addition to the matrices $X$ and $L$, which can be obtained by solving two Lyapunov equations~\eqref{eq.lyap1} and~\eqref{eq.ExtraLyap}, respectively, the computation of the derivative of the cost function ($J$) requires the derivatives $\partial \Gamma_{a_\mathrm{m}}/\partial w_i$, which can be expressed analytically. The dependence of the augmentation operator $\Gamma_{a_\mathrm{m}}$ on the weights $w_i$ can be decomposed into two contributions: one mediated by the streamwise mean ($\bar{U}$), and another arising from the direct dependence on the Fourier coefficients $a_m$, which themselves depend on $w_i$ through Eq.~\eqref{eq.SDFVariableChange}. Specifically, 
\begin{align}
    \Gamma_{a_\mathrm{m}}
    \;=\;
    \Gamma_{\text{mean}}\left(\bar{U}\left(a_\mathrm{m} \right) \right)
    \,+\,
    \Gamma_{\text{fluc}}\left(a_\mathrm{m}\right).
\label{sumFeedbackDyn}
\end{align} 
Since scalar differentiation commutes with linear and diagonal operators, we obtain
 \begin{align*}
 \ba{rcl}
 \dfrac{\partial }{\partial w_i}\left(\Gamma_{\text{mean}}\left(\bar{U}\left(a_\mathrm{m} \right)\right)\right)
 &\!\!=\!\!&
 \Gamma_{\text{mean}}\left(\dfrac{\partial \bar{U}\left(a_\mathrm{m}\right)}{\partial w_i} \right)
 \\[.35cm]
 \dfrac{\partial }{\partial w_i}\left(\Gamma_{\text{fluc}}\left(a_\mathrm{m}\right)\right)
 &\!\!=\!\!&
 \Gamma_{\text{fluc}}\left(\dfrac{\partial a_\mathrm{m}}{\partial w_i} \right).
 \ea
 \end{align*}
Thus, computing $\partial \Gamma_{\mathbf w}/\partial w_i$
reduces to evaluating ${\partial a_\mathrm{m}}/{\partial w_i}$ and ${\partial \bar{U}\left(a_\mathrm{m}\right)}/{\partial w_i}$. 

To compute ${\partial a_\mathrm{m}}/{\partial w_i}$, we exploit the linearity of the Fourier transform in the expression for $a_\mathrm{m}$, which allows the derivative to be moved inside the transform. Applying the chain rule yields 
\begin{align*}
\dfrac{\partial a_\mathrm{m}}{\partial w_i}
\;=\;
\mathcal{FT}\left\{ \dfrac{\phi_i e^{-\sum \phi_i w_i / \epsilon}}{\epsilon\left(1+e^{-\sum \phi_i w_i / \epsilon}\right)^2} \right\}.
\end{align*}
To compute ${\partial \bar{U}\left(a_\mathrm{m}\right)}/{\partial w_i}$, we first express the mean $\bar{U}$, which solves Eq.~\eqref{meanRef}, as
\begin{align*}
\bar{U} \;=\; {\cal A}_{\mathrm{mean}}^{-1}\, {\cal F}_P
\end{align*}
where ${\cal A}_\mathrm{mean}\DefinedAs {\cal A}_\text{s} + {\cal A}_\text{rib}(a_\mathrm{m})$. Here, ${\cal A}_\text{s}$ and ${\cal A}_\text{rib}$ denote the dynamical contributions from the smooth wall and the riblets, respectively, and ${\cal F}_P$ is the forcing vector associated with the nominal back pressure. Using the derivative of an inverse operator, together with the linearity of ${\cal A}_{\text{rib}}$, gives the closed-form expression
\begin{align*}
    \dfrac{ \partial \bar{U}(a_\mathrm{m})}{\partial w_i}
    \;=\;-{\cal A}_\mathrm{mean}^{-1} \,{\cal A}_{\text{rib}}\!\left(\dfrac{ \partial a_{\mathrm{m}}}{\partial w_i}\right) {\cal A}_\mathrm{mean}^{-1}\, {\mathcal F}_P.
\end{align*}

\section*{Acknowledgments}
The Office of Information Technology Cyberinfrastructure Research Computing (CIRC) at The University of Texas at Dallas and the Texas Advanced Computing Center (TACC) are acknowledged for providing computing resources. Modern LLM's were utilized in accordance with the IEEE GenAI guidelines for brainstorming, reference identification, and grammatical editing for enhanced textual clarity.

\end{document}